\documentclass[sigconf, authorversion, nonacm]{acmart}

\AtBeginDocument{%
  }

\setcopyright{acmlicensed}
\copyrightyear{2018}
\acmYear{2018}
\acmDOI{XXXXXXX.XXXXXXX}
\acmConference[Conference acronym 'XX]{Make sure to enter the correct
  conference title from your rights confirmation email}{June 03--05,
  2018}{Woodstock, NY}
\acmISBN{978-1-4503-XXXX-X/2018/06}

\usepackage{xspace}
\usepackage{algpseudocode}
\usepackage{algorithm}
\usepackage{multirow}
\usepackage[inline]{enumitem}
\usepackage{wrapfig}
\usepackage{subcaption}
\usepackage[export]{adjustbox} 
\usepackage{tcolorbox}
\usepackage{listings}
\usepackage{xcolor}
\colorlet{sigpositive}{blue}
\colorlet{signegative}{orange}
\usepackage{booktabs}
\usepackage{cleveref}

\newtcolorbox{promptbox}[1]{
    colback=white,
    colframe=black!70,
    fonttitle=\bfseries,
    title=#1,
    arc=0pt,
    outer arc=0pt,
    left=8pt,
    right=8pt,
    top=8pt,
    bottom=8pt,
    fontupper=\small\ttfamily
}

\usepackage{xcolor}

\usepackage{booktabs}
\usepackage{longtable}
\usepackage{array}
\usepackage{ragged2e}

\usepackage{listings}

\lstdefinestyle{llmprompt}{
basicstyle=\ttfamily\footnotesize,
commentstyle=\ttfamily\footnotesize,
keywordstyle=\ttfamily\footnotesize,
stringstyle=\ttfamily\footnotesize,
breaklines=true,
columns=fullflexible,
keepspaces=true,
showstringspaces=false,
frame=single,
xleftmargin=2pt,
xrightmargin=2pt
}

\begin{document}

%\title{Scaffold or Co-Write? Comparing AI Support Modalities in Gratitude Writing}
\title{Supporting Without Taking Over: AI Support in Gratitude Writing}

\author{Thu Bui*}
\email{bui35@purdue.edu}
\orcid{XXXX-XXXX-XXXX}
\affiliation{%
  \institution{Department of Computer Science\\Purdue University}
  \city{West Lafayette}
  \state{IN}
  \country{USA}
}

\author{Shashi Salavath*}
\email{ssalavat@purdue.edu}
\orcid{XXXX-XXXX-XXXX}
\affiliation{%
  \institution{Department of Computer Science\\Purdue University}
  \city{West Lafayette}
  \state{IN}
  \country{USA}
}

\author{Hillary Merzdorf}
\email{hmerzdor@purdue.edu}
\orcid{XXXX-XXXX-XXXX}
\affiliation{%
  \institution{Department of Psychological Sciences\\Purdue University}
  \city{West Lafayette}
  \state{IN}
  \country{USA}
}

\author{Louis Tay}
\email{stay@purdue.edu}
\orcid{XXXX-XXXX-XXXX}
\affiliation{%
  \institution{Department of Psychological Sciences\\Purdue University}
  \city{West Lafayette}
  \state{IN}
  \country{USA}
}

\author{Sooyeon Jeong}
\email{sooyeonj@purdue.edu}
\orcid{XXXX-XXXX-XXXX}
\affiliation{%
  \institution{Department of Computer Science\\Purdue University}
  \city{West Lafayette}
  \state{IN}
  \country{USA}
}

\begin{abstract}

Gratitude writing can support well-being, but independent practice may lack guidance. Generative AI could help while shaping reflection, agency, and ownership. We investigated how gratitude-writing format and AI-assistance mode influence well-being and writing experiences. In a 2$\times$2 between-subjects study ($N=401$), participants completed a \emph{gratitude reflection} or a \emph{gratitude letter} using two AI-assistance modes: \emph{scaffolding}, which provided reflective questions after inactivity, or \emph{co-writing}, which offered inline text suggestions. Participants' well-being improved after using our AI-assisted gratitude-writing system, consistent with the activity's intended benefits. However, participants' experiences revealed a trade-off between the AI-support types: co-writing received higher usability ratings, yet its immediacy could disrupt writing, and greater suggestion acceptance was associated with weaker perceived agency. Scaffolding supported stronger perceived human ownership but was sometimes overlooked. These findings extend human–AI writing research beyond productivity and text quality by showing how AI involvement shapes psychological outcomes and the experience of self-guided, emotionally meaningful writing.
\end{abstract}




\received{20 February 2007}
\received[revised]{12 March 2009}
\received[accepted]{5 June 2009}

\maketitle
\begin{figure*}[t]
\centering
\includegraphics[width=\textwidth]{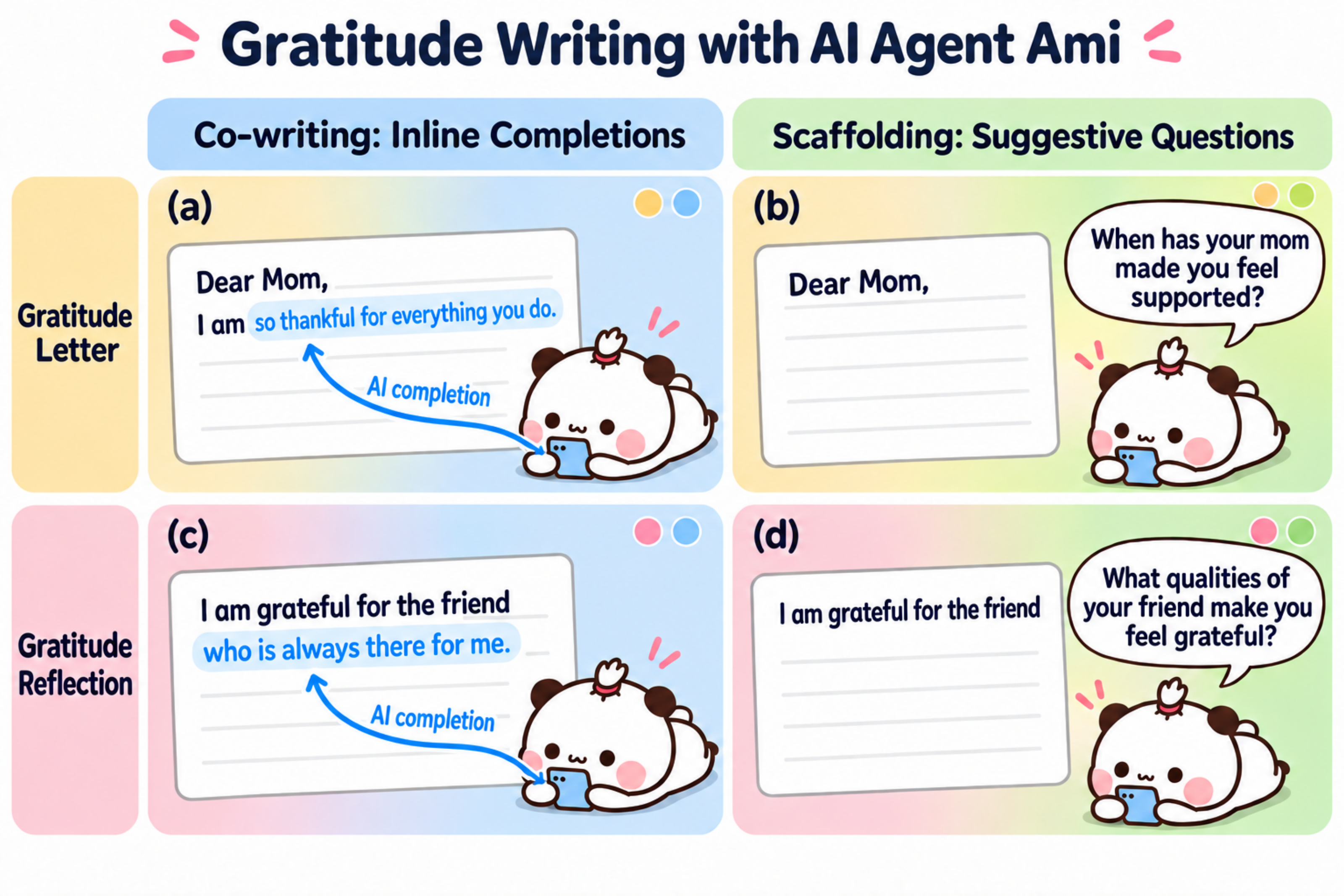}
\caption{Illustrative overview of our $2\times2$ study design, combining two gratitude-writing formats (rows) with two AI-assistance modes (columns). Participants either wrote a gratitude letter (a,b) to a specific person or a gratitude reflection (c,d) on multiple sources of gratitude in their lives. (a,c) In co-writing, Ami offers inline text completions for gratitude letters and reflections. (b,d) In scaffolding, Ami poses open-ended reflective questions without contributing language directly to the document.}
\Description{A four-panel cartoon matrix titled `Gratitude Writing with AI Agent Ami.'' The top row represents gratitude-letter writing, and the bottom row represents gratitude reflection. The left column shows co-writing: Ami provides inline completions for `Dear Mom'' in panel a and ``I am grateful for the friend'' in panel c. The right column shows scaffolding: Ami asks when the writer's mother made them feel supported in panel b and which qualities of a friend inspire gratitude in panel d. Ami appears beside the writing page in each panel.}
\label{fig:study-teaser}
\end{figure*}

\section{Introduction}

Generative language models have expanded the role of AI in writing from offering feedback and suggestions to generating ideas, sentences, and longer passages alongside writers~\cite{lee2022coauthor}. Although these systems can help writers initiate and develop text, their involvement can also shape what people express and how they perceive their contribution to the resulting text~\cite{jakesch2023co,draxler2024ai}. These implications may be especially consequential when writing is undertaken not only to produce an artifact but also to support psychological well-being. AI assistance could help people identify and articulate personally meaningful experiences; however, directly generating their words might alter the reflection, effort, and sense of authorship that structured writing activities produce~\cite {frattaroli2006experimental}. This tension is particularly relevant to expressive writing interventions~\cite{frattaroli2006experimental,pennebaker1997writing}, which use the structured articulation and reflection of personal experiences to support mental health and psychological well-being.

Within the broader landscape of well-being-oriented writing, gratitude writing directs attention toward positive experiences, meaningful relationships, and sources of appreciation~\cite{wood2010gratitude,davis2016thankful,diniz2023effects}. By helping individuals notice, elaborate on, and assign meaning to positive aspects of their lives, gratitude writing interventions can foster positive affect, strengthen feelings of appreciation, and improve subjective well-being~\cite{emmons2003counting,davis2016thankful,wood2010gratitude}. These interventions can structure gratitude in different ways: some ask individuals to reflect privately on valued experiences, whereas others ask them to formulate and communicate their appreciation through a gratitude letter~\cite{davis2016thankful,diniz2023effects}. Despite their general benefits, gratitude writing interventions are not uniformly effective; their outcomes may depend on factors such as intervention format, duration, motivation, and effort~\cite{sin2009enhancing,lyubomirsky2011becoming}. How an intervention structures reflection and expression may therefore shape both its well-being benefits and the writer’s experience, particularly as many people nowadays use generative AI to generate ideas or language.

In writing tasks, generative AI can assist people through different levels of involvement. In a scaffolding role, AI can help writers generate and explore topics, questions, or possible directions while leaving the composition of the text to the writer. In a co-writing role, AI can contribute language directly and take a more active role in determining how ideas are articulated in the final text. Prior research demonstrates that language models can support both ideation and text production~\cite{lee2022coauthor}. However, AI assistance can also shape what writers express and create a divergence between their perceived ownership of AI-generated text and their claims of authorship~\cite{jakesch2023co,draxler2024ai}. These distinctions may be especially consequential in well-being interventions that involve writing tasks, where the process of reflection may matter as much as, or more than, the production of polished text. Scaffolding assistance could facilitate reflection while preserving the writer’s direct contribution, whereas co-writing could reduce barriers to expression while also altering effort, agency, or perceived authorship. Yet little is known about how these modes of AI assistance affect the psychological outcomes and subjective experience of gratitude writing.

These writing formats and assistance modes may also shape what personal information writers choose to disclose. Private gratitude reflection and a gratitude letter frame expression differently. One emphasizes examining personal experiences, while the other asks writers to articulate appreciation to a recipient. Because both reflective questions and suggested wording can introduce directions for personal expression, AI assistance may influence what writers choose to share. We therefore examine whether self-disclosure differs across scaffolding and co-writing conditions. Examining self-disclosure helps us understand how these design choices shape the personal content of gratitude writing alongside its psychological outcomes and perceived authorship.

To investigate these questions, we conducted a 2×2 between-subjects study that varied \textbf{gratitude-writing format} (\textit{gratitude reflection} vs. \textit{gratitude letter}) and \textbf{AI-assistance mode} (\textit{scaffolding} vs. \textit{co-writing}), as illustrated in \Cref{fig:study-teaser}. We aimed to investigate whether the effects of AI assistance depend on the type of gratitude exercise in which it is embedded. We examined this interaction across participants’ well-being outcomes and perceptions of the writing experience. Specifically, we asked:

\begin{itemize}
    \item \textbf{RQ1}. How do gratitude-writing type and AI-support type shape participants' well-being outcomes?
    \item \textbf{RQ2}. How do gratitude-writing type and AI-support type shape participants' perceptions of the writing experience, including efficacy, ownership, control, leadership, and their experience of interacting with the AI?
    \item \textbf{RQ3}. How do gratitude-writing type and AI-support type shape the extent and nature of personal disclosure in gratitude writing?
\end{itemize}

Our findings reveal distinct benefits and trade-offs across gratitude-writing formats and AI-assistance modes. Most well-being measures improved from pre- to post-writing across all four conditions, although the pattern of gains differed by outcome. Gratitude reflection produced larger gains in life satisfaction and more thought-based disclosure, whereas gratitude letters elicited more feelings-based disclosure. Co-writing produced larger gains in positive affect and affect balance and received higher user-experience and usability ratings, while scaffolding better preserved perceived human ownership. AI-assistance mode did not significantly affect LLM-rated self-disclosure; differences in disclosure were observed between writing formats.

Interaction behavior further revealed a usability-agency trade-off. Within co-writing, greater suggestion uptake is associated with higher user-experience ratings but lower perceived ownership, control, and human attribution. Across both assistance modes, engagement was more consistently associated with interface experience than with well-being change. Qualitative findings contextualize this trade-off: participants valued both systems for developing ideas and overcoming writing difficulties. Co-writing offered immediately usable language but sometimes interrupted the writing process, while scaffolding encouraged reflection in participants' own words but was sometimes difficult to notice. These findings highlight complementary benefits and trade-offs of AI assistance in well-being-oriented writing and motivate future exploration of support that dynamically adapts to writers' needs in real time while preserving their control over AI involvement.

\section{Related Works}
\subsection{Writing and Gratitude Interventions for Well-being}
Structured writing has long been studied as a means of supporting psychological and physical well-being. Early expressive writing research asked individuals to write about stressful, traumatic, or emotionally significant experiences, often across several brief sessions~\cite{pennebaker1997writing}. Rather than focusing on the quality of the resulting text, these interventions treat writing as a process through which people articulate emotions, organize experiences, and construct meaning. Meta-analytic evidence indicates that expressive writing can produce small but reliable benefits across psychological, behavioral, and physical-health outcomes, although these effects vary across studies and participants~\cite{frattaroli2006experimental}. This variability suggests that the benefits of writing depend not simply on whether people write, but on what they write about and how they engage in the activity.

Subsequent research expanded writing interventions beyond the disclosure of difficult experiences to include reflection on positive experiences and possible futures. For instance, writing about intensely positive experiences has been associated with improvements in health and well-being~\cite{burton2004health}. Other positive writing activities ask participants to imagine their best possible selves \cite{king2001health}, identify positive daily events \cite{seligman2005positive}, or reflect on sources of meaning and appreciation \cite{emmons2003counting}. Although these interventions vary in content and structure, they share the premise that directing and elaborating attention through writing can influence psychological well-being. Positive activity models~\cite{lyubomirsky2011becoming,lyubomirsky2013simple} further propose that their effectiveness depends on characteristics of both the activity and the participant, including dosage, variety, motivation, effort, and person–activity fit.

Gratitude writing is one widely studied form of positive writing intervention. Emmons and McCullough~\cite{emmons2003counting} found that participants who regularly recorded things for which they were grateful reported greater well-being and optimism—and, in some studies, fewer physical symptoms—than participants who recorded hassles or neutral events. Seligman et al.~\cite{seligman2005positive} similarly found that gratitude-focused activities could increase happiness and reduce depressive symptoms. Meta-analyses have since identified reliable benefits of gratitude interventions while documenting variation across outcomes, comparison conditions, intervention durations, and activity formats~\cite{davis2016thankful,dickens2017using,diniz2023effects}. Gratitude writing may produce these benefits by directing attention toward positive experiences and meaningful relationships and encouraging people to elaborate on why they matter.

Gratitude interventions structure reflection and expression in different ways. Individually oriented exercises, often described as \emph{gratitude listing} or \emph{counting blessings}, ask participants to identify and elaborate on positive experiences, people, or circumstances for which they feel grateful~\cite{emmons2003counting,froh2008counting}. Interpersonally oriented exercises instead ask participants to compose a \emph{gratitude letter} expressing appreciation to a specific benefactor~\cite{seligman2005positive,toepfer2012letters}. Whereas the former centers private reflection, the latter requires writers to formulate their gratitude in relation to another person and a potential audience. 

Meta-analytic findings support a small but reliable overall effect of gratitude interventions on well-being while also indicating variation across intervention formats and study designs~\cite{davis2016thankful,dickens2017using}. Specifically, both reflective and letter-writing activities were found effective in promoting positive outcomes, but their effects may differ. Watkins et al.~\cite{watkins2003gratitude} found that thinking about, writing about, and composing a letter to a person for whom one felt grateful all increased immediate positive affect relative to a neutral task, while repeated gratitude-letter writing has been associated with cumulative improvements in happiness and life satisfaction~\cite{toepfer2012letters}. Prior work additionally suggests that gratitude intervention formats can differ in their perceived difficulty and likelihood of completion~\cite{kaczmarek2015people}. These differences motivate our comparison of gratitude reflection and gratitude-letter writing.

\subsection{Generative AI Assistance for Writing}
In conventional gratitude interventions, instructions structure the topic and form of reflection, but participants remain responsible for generating and articulating their responses. However, recent advances in generative AI can potentially change this by allowing an interactive AI system to collaborate with users in generating ideas, composing language, or both. This raises new questions about how AI involvement interacts with the distinct reflective and interpersonal demands of gratitude-writing activities.

A review of 115 intelligent writing-assistant studies identified five dimensions along which such systems vary: task, user, technology, interaction, and ecosystem~\cite{lee2024designspace}. Of particular relevance to our work is the nature of generative AI's contribution. Some systems support idea generation and exploration by offering conceptual prompts, questions, potential topics, or alternative directions during prewriting~\cite{gero2022sparks,wan2024felt,gero2023social}, whereas others generate continuations, rewrites, elaborations, or complete passages that writers can incorporate directly into their text~\cite{yuan2022wordcraft,lee2022coauthor,chen2019smart}. 

These roles distribute the reflective and compositional work differently between the writer and the system. In direct co-writing systems, writers and AI alternate or combine their contributions to produce a text. Systems, such as Wordcraft~\cite{yuan2022wordcraft} and CoAuthor~\cite{lee2022coauthor}, allow writers to request continuations, elaborations, and rewrites at different points in the composition process. Wordcraft~\cite{yuan2022wordcraft} participants used language-model assistance to overcome blocks and develop story details, although the system sometimes struggled to preserve writers' intentions and authorial voices. CoAuthor~\cite{lee2022coauthor} captured 1,445 creative and argumentative writing sessions in which participants could request, accept, edit, or reject GPT-3 suggestions. Writers accepted 72.3\% of the suggestions they requested, yet humans still produced 72.6\% of the final text, which illustrated that writers selectively incorporated AI-generated language rather than simply delegating the entire task.

Other systems position AI primarily as a source of ideas. Sparks~\cite{gero2022sparks}, for example, generated conceptually related prompts intended to inspire science writers rather than provide polished prose for direct insertion. Participants used these prompts to find inspiration, translate technical concepts, and consider alternative perspectives. Related research on predictive and parallel suggestions demonstrates that AI-generated options can influence both what writers consider and what they ultimately produce~\cite{arnold2020predictive,buschek2021impact}. Idea-oriented assistance may therefore expand the possibilities available to writers while preserving their responsibility for selecting, interpreting, and articulating those ideas. In contrast, direct co-writing shifts part of the compositional work to the system.

Greater AI involvement can make writing easier, but it may also affect agency, ownership, and the diversity of written expression. Predictive suggestions can steer writers toward language made salient by the system~\cite{arnold2020predictive}, while AI-generated contributions can create discrepancies between users’ perceived ownership of a text and their attribution of authorship~\cite{draxler2024ai}. Other studies suggest that reliance on language-model suggestions can reduce linguistic or content diversity and shift writers from generating ideas toward evaluating AI-provided options~\cite{padmakumar2024diversity,agarwal2025homogenize,bhat2026reactive}. The degree of assistance also appears consequential: different levels of generated support can produce tradeoffs among writing quality, cognitive investment, and perceived ownership~\cite{dhillon2024shaping}. Together, these findings show that AI assistance changes not only the efficiency or quality of writing but also how writers participate in producing the text.

These consequences may be particularly important in well-being-oriented writing, where reflection and personal articulation are part of the intended activity rather than merely steps toward a finished artifact. Scaffolding assistance may help people identify and develop meaningful experiences while leaving them responsible for expressing those experiences. Co-writing may reduce barriers to expression by contributing language directly, but it may also shift effort, agency, or ownership away from the writer. Prior work has rarely examined whether these modes of assistance affect the psychological outcomes of writing interventions or whether their effects depend on the reflective and interpersonal demands of the intervention itself.

\section{Method}
\subsection{Study Design}
We conducted an online \(2 \times 2\) between-subjects study to examine how \textbf{gratitude intervention format} and \textbf{AI-assistance mode} jointly influence the experience and outcomes of gratitude writing. Gratitude intervention format consisted of \textit{gratitude reflection} and \textit{gratitude letter writing}, while AI-assistance mode consisted of \textit{scaffolding support} and \textit{co-writing support}. Upon informed consent, participants were randomly assigned to one of four conditions: (1) {gratitude reflection with scaffolding (RS)}, (2) {gratitude reflection with co-writing (RC)}, (3) {gratitude letter with scaffolding (LS)}, and (4) {gratitude letter with co-writing (LC)}. All conditions followed the same study procedure and used a shared interface, condition-specific writing instruction and AI-assistance features described below. 

\subsubsection{Gratitude-Writing Format}
The gratitude-writing activities differed in their focus and intended audience. In the \textit{gratitude reflection} condition, participants were asked to write about positive experiences, meaningful relationships, or other aspects of their lives for which they felt grateful \cite{emmons2003counting,regan2023some}. In the \textit{gratitude letter} condition, participants addressed their writing to a specific person and explained why they appreciated that person \cite{regan2023some,seligman2005positive}. \textit{Gratitude reflection} therefore allowed participants to explore different sources of gratitude, whereas \textit{gratitude letter} writing focused expression on a particular recipient. The complete writing instructions are provided in \Cref{app:writing_prompt}.

\subsubsection{AI-assistance Modes}

Both AI-assistance modes used GPT-4o with condition-specific prompts that incorporated the assigned writing task and the participants' current written draft. The modes differed in the content, presentation, and timing of the AI assistance. 
In scaffolding, the AI agent, referred to as ``Ami'' in the web interface, supported idea development through open-ended questions, topics, and possible directions. For example, it could ask ``How has your friendship influenced or enhanced your experience with your hobby?'' Scaffolding appeared when requested by participants or automatically after 20 seconds of writing inactivity. Participants could provide feedback on each question using thumbs-up or thumbs-down buttons and remained responsible for composing the text in their own words.

In co-writing, the AI-support feature functioned as an autocomplete tool that suggested words or sentences to continue the participant's written draft based on the text before the cursor. 
For example, after a participant typed ``\textit{Dear Baba, I just wanted to start off by saying}'', the system suggested ``\textit{thank you for always being there for me.}'' Inline suggestions appeared after approximately one second of writing inactivity. Participants accept a suggestion by pressing the \textit{Tab} key or reject/ignore the suggestion and continue writing their own text. Accepted text could be edited after insertion. Co-writing also provided side-panel guidance through the \emph{Ask Ami} button. This assistance appeared only when requested and provided a direct writing suggestion. For example, when the writing contained ``\textit{Dear Mom,}'' the system provided direct writing guidance rather than a reflective question: ``Include Specific Memories or Actions: To make your letter more heartfelt, recall specific moments or actions that made you grateful for your mom. For example, you might write about a time she supported you during a challenging situation or a cherished family tradition she started. These details can make your gratitude more tangible and personal.'' 

\begin{figure*}[t]
\centering
\includegraphics[width=\textwidth]{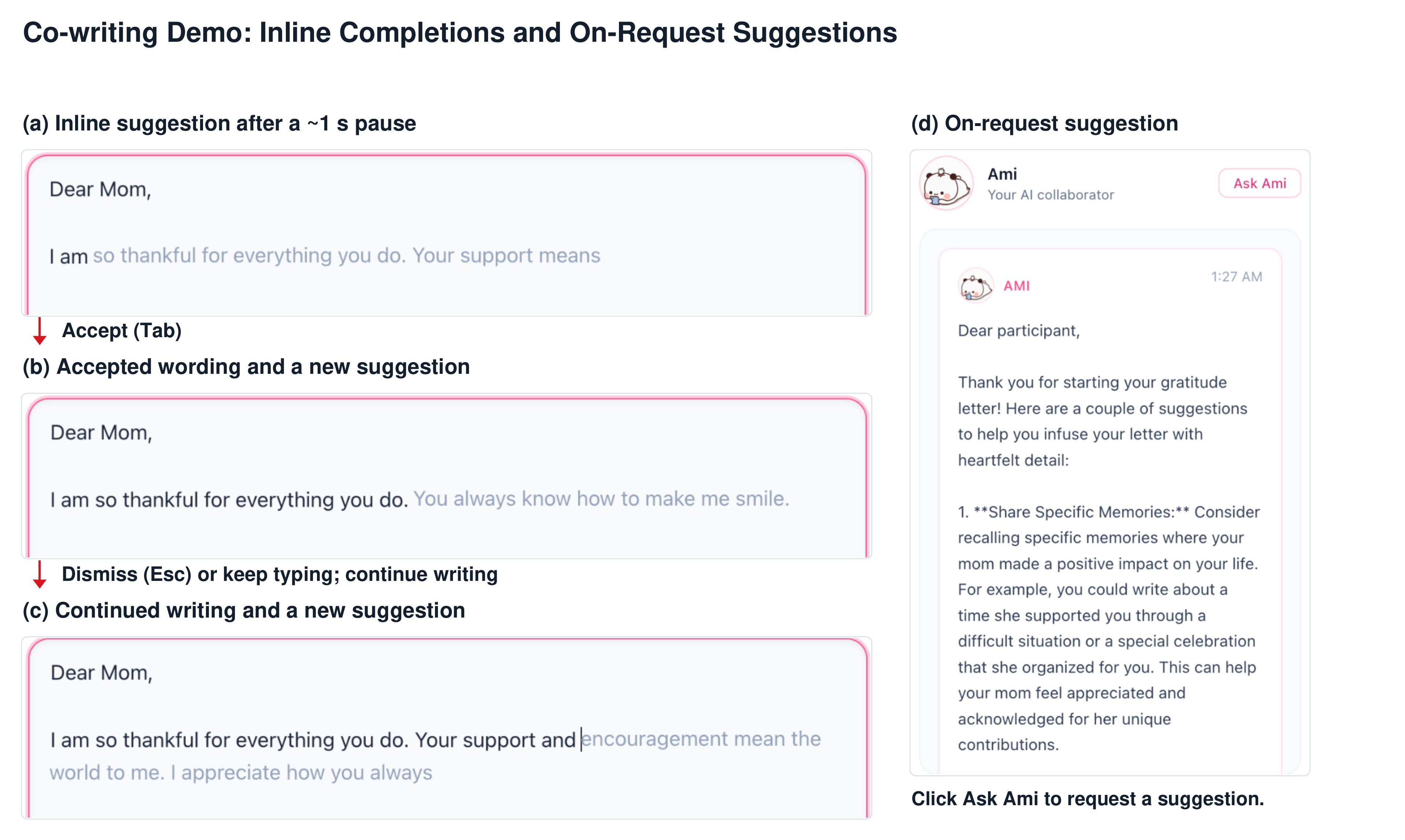}
\caption{Detailed interaction flow for co-writing. (a) After approximately one second of inactivity, the system displays a pending inline completion in faded text. (b) Pressing Tab accepts and incorporates the suggestion into the document, after which another completion may appear following a subsequent one-second pause. (c) Pressing Esc or continuing to type dismisses or ignores the pending completion; the system may offer a new suggestion after the writer pauses again. (d) Participants can also select ``Ask Ami'' to request broader writing suggestions in the side panel. Co-writing therefore provides both immediate sentence-level completions and participant-initiated writing support.}
\Description{A four-panel co-writing demonstration. Three vertically arranged panels on the left show a gratitude letter receiving a faded inline completion, the completion becoming dark text after acceptance, and the writer continuing after dismissing a later completion. Arrows label the Tab acceptance and Esc dismissal actions. A panel on the right shows an Ask Ami button and a longer, participant-requested suggestion in the side panel.}
\label{fig:cowriting_demo}
\end{figure*}

\begin{figure*}[t]
\centering
\includegraphics[width=\textwidth]{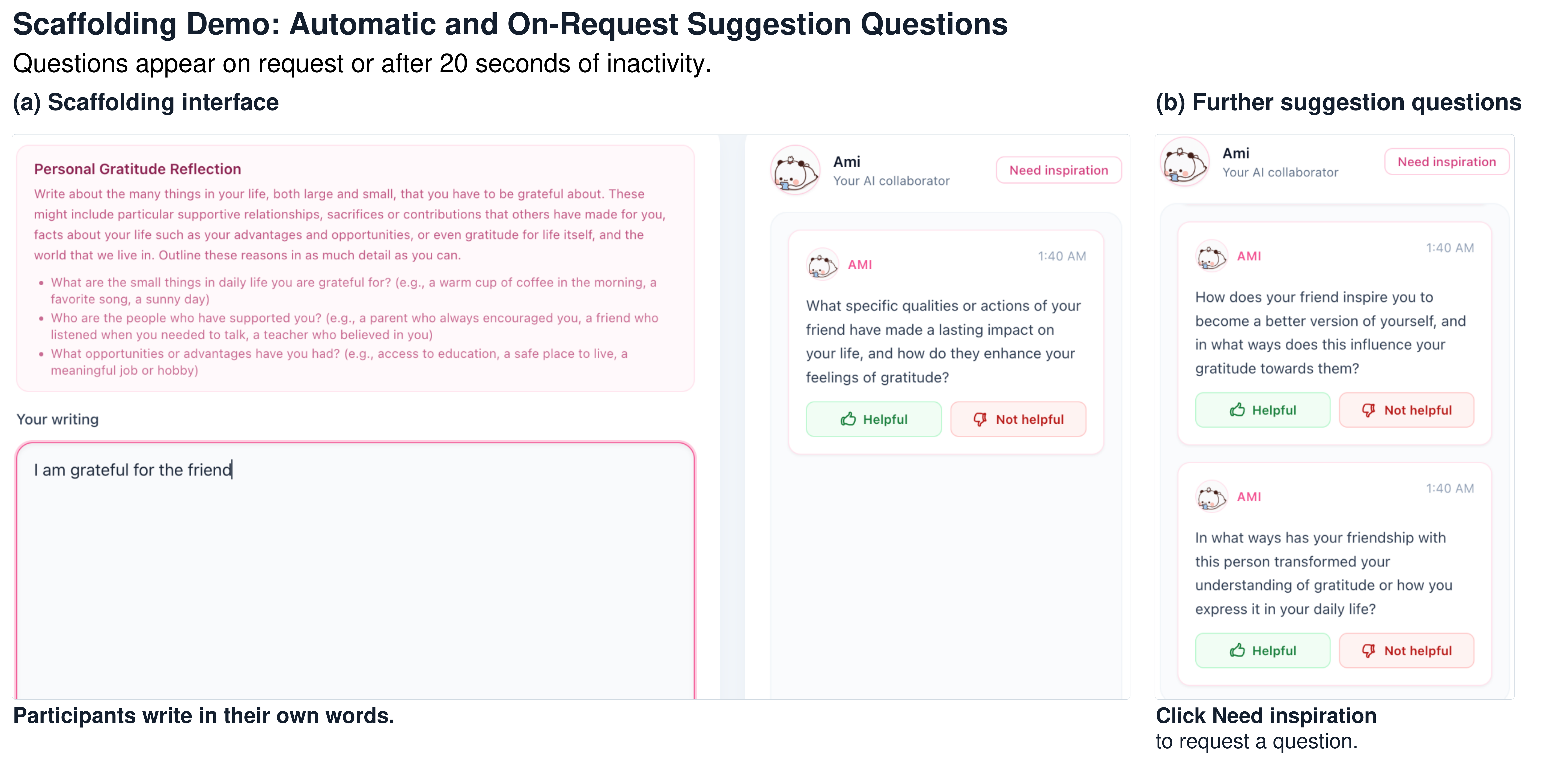}
\caption{Detailed interaction flow for scaffolding. (a) As participants compose the document in their own words, the system presents an open-ended reflective question in the side panel, either automatically after 20 seconds of inactivity or at the participant's request. Participants can rate each question as helpful or not helpful. (b) Selecting ``Need inspiration'' immediately generates an additional question; repeated requests can produce different prompts even when the underlying text remains unchanged. Scaffolding thereby supports continued reflection without contributing language directly to the document.}
\Description{A two-panel scaffolding demonstration. The larger panel on the left shows a gratitude-reflection writing area containing participant-authored text and a side panel containing one reflective question with Helpful and Not helpful buttons. The narrower panel on the right shows two additional reflective questions in the side panel and the Need inspiration button used to request further guidance.}
\label{fig:scaffolding_demo}
\end{figure*}

The comparison therefore examined two configurations of AI assistance that differed in both their contribution to the text and their interaction design. The complete model prompts are provided in \Cref{app:llm_scaffolding,app:llm_cowriting}. \Cref{fig:cowriting_demo,fig:scaffolding_demo} illustrate the co-writing and scaffolding interaction sequences, respectively. Supplementary videos demonstrate all four experimental conditions.

\subsection{Participants}
We recruited participants through the Purdue's research participant pool and Prolific\footnote{https://www.prolific.com/}. To be eligible from the university's research pool, student participants were required to be fluent in English and enrolled as a sophomore, junior, senior, or graduate student. They received two research credits for participating in the study. Prolific participants were required to be at least 18 years old, fluent in English, and have access to a computer. They received $\$6$ for completing the study. The study protocol was reviewed and approved by the Institutional Review Board, and all participants provided informed consent before beginning the study.

A total of 612 individuals began the study according to our online database. Of these, 496 could be linked to a writing-session record in our study database. Before conducting the primary analyses, we excluded 89 participants who submitted no written response, 3 who did not comply with the study procedure (e.g., copying/pasting text generated outside the assigned study interface), and 3 whose interaction records were incomplete because of technical errors. The final analytic sample consisted of 401 participants: 372 recruited through Prolific and 29 through the Purdue University's online student participant pool.

The analytic sample includes 401 participants: 104 in \textit{gratitude-letter}/\textit{co-writing} (LC), 107 in \textit{gratitude-letter}/\textit{scaffolding} (LS), 98 in \textit{gratitude reflection}/\textit{co-writing} (RC), and 92 in \textit{gratitude reflection}/\textit{scaffolding} (RS). Because survey responses were optional, missing responses resulted in different sample sizes across measures. The number of participants included in each measure analysis is reported in \Cref{app:numpartcipant}.

\subsection{Writing Interface}
\subsubsection{Interface Layout and Features}

\begin{figure*}[t]
\centering
\includegraphics[width=\textwidth]{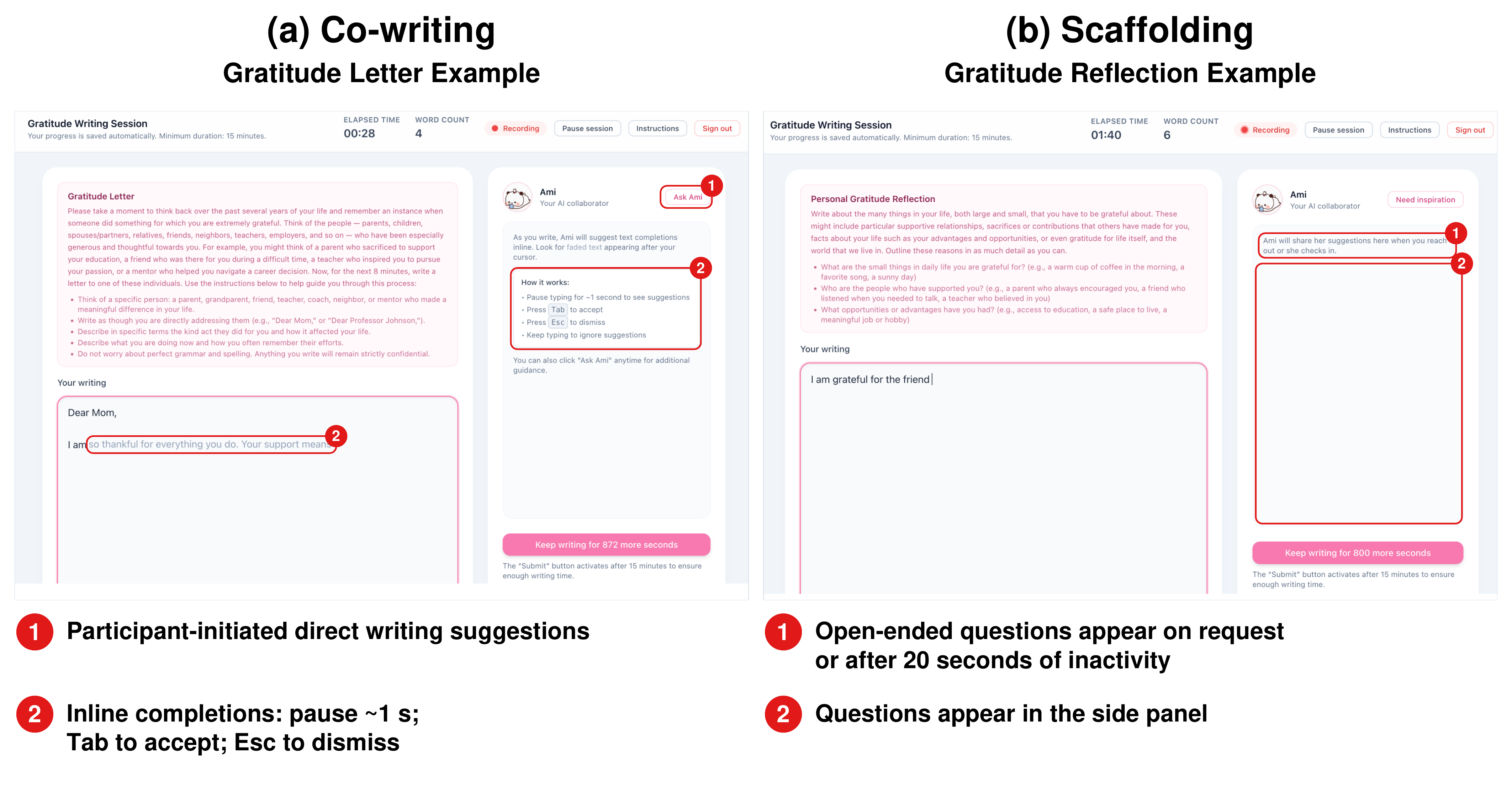}
\caption{Overview of the two AI-support interfaces, illustrated with a gratitude-letter example with co-writing and a gratitude-reflection example with scaffolding. (a) In co-writing, participants can request writing suggestions through ``Ask Ami.'', with suggestions displayed in the side panel. The system also displays inline completions after approximately one second of inactivity; participants can accept a completion with Tab, dismiss it with Esc, or ignore it by continuing to type. (b) In scaffolding, open-ended reflective questions appear in the side panel either when participants select ``Need inspiration'' or automatically after 20 seconds of inactivity.}
\Description{Two side-by-side screenshots compare the AI-support interfaces. The left screenshot shows a gratitude-letter co-writing interface with a writing area containing a faded inline completion and a side panel containing an Ask Ami button and instructions for accepting or dismissing suggestions. The right screenshot shows a gratitude-reflection scaffolding interface with a writing area and an empty side panel where reflective questions will appear. Numbered callouts identify the request buttons and the locations and interaction rules for each form of assistance.}
\label{fig:system-interface}
\end{figure*}

Participants completed the gratitude-writing task through a custom web interface embedded in Qualtrics. As shown in \Cref{fig:system-interface}, the interface used the same overall layout across conditions: the assigned writing instructions appeared above a large text editor on the left, while an AI-support panel featuring the agent, Ami, appeared on the right. The side panel explained the available AI interaction, displayed generated guidance, and included a countdown indicating the remaining required writing time. Participants' drafts were saved automatically, a word count was displayed below the editor, and the submission button became available after 15 minutes. The primary differences across conditions were the writing instructions and the way participants interacted with Ami.

In co-writing, Ami displayed candidate continuations as faded inline text after participants paused for approximately one second. Participants could press \texttt{Tab} to accept a completion, press \texttt{Esc} to dismiss it, or continue typing to ignore it. They could also click the \emph{Ask Ami} button to request an additional suggestion in the side panel. In scaffolding, Ami did not insert language into the document. Instead, after 20 seconds of inactivity, the side panel automatically displayed a reflective question intended to help participants develop their ideas. Participants could rate each question as helpful or not helpful and could also request support through the \emph{Need inspiration} button.

The shorter delay in co-writing was intended to provide immediate assistance while maintaining writing momentum. The longer delay in scaffolding treated sustained inactivity as a possible indication that the writer needed guidance, allowing the system to intervene without repeatedly interrupting active writing.

\subsubsection{Interaction Logging}
The system captured a recording of the study browser tab and logged participants' writing and interactions with Ami. For co-writing, the logs recorded each inline suggestion displayed and whether it was accepted, dismissed, or ignored, as well as requests submitted through \emph{Ask Ami}. A suggestion was considered accepted when the participant inserted it using \texttt{Tab}; suggestions dismissed using \texttt{Esc} or superseded by continued typing were treated as rejected. We calculated the acceptance rate as the number of accepted suggestions divided by the number displayed, and the resolved acceptance rate as the number accepted divided by the total number accepted or rejected.

For scaffolding, the logs recorded each question displayed and participants' helpful or not-helpful ratings. The like rate was calculated as the number of helpful ratings divided by the total number of helpful and not-helpful ratings among participants who rated at least one prompt. The system also saved participants' final submitted texts for subsequent analysis. 

\subsection{Study Procedure}
Recruited participants were directed to a Qualtrics survey, where they reviewed the study information, provided informed consent, and completed baseline measures of life satisfaction, trait gratitude, positive and negative affect, and moral elevation (see \Cref{sec:measures,app:measures}). Qualtrics then presented the embedded writing interface, which delivered the writing task and AI support corresponding to each participant's assigned condition. Participants completed the writing activity for at least 15 minutes, during which their writing and system interactions were recorded as described above.

After submitting their writing, participants returned to Qualtrics and completed the same well-being measures, followed by measures of writing self-efficacy, perceived ownership, control, leadership, text match, user experience, and usability. They also answered four open-ended questions about the system and their writing experience. Upon completion, Prolific participants received \$6, while participants recruited through the university's research participant pool received two research credits.

\subsection{Measures}
\label{sec:measures}
\subsubsection{Self-reported Measures}
We assess four dimensions of well-being before and after the writing intervention, with measures selected based on prior work \cite{diniz2023effects}. The five-item Satisfaction With Life Scale (SWLS) measures participants' global cognitive evaluation of their life satisfaction \cite{diener1985satisfaction}. The Gratitude Questionnaire--Six-Item Form (GQ-6) assesses the dispositional tendency to experience and express gratitude \cite{mccullough2002grateful}. The 12-item Scale of Positive and Negative Experience (SPANE) measures the frequency of positive and negative feelings \cite{diener2010new}, producing separate positive-affect (SPANE-P) and negative-affect (SPANE-N) scores, as well as an affect-balance score calculated as SPANE-B = SPANE-P $-$ SPANE-N. The Moral Elevation measure captures feelings of being moved, uplifted, or inspired by witnessing morally admirable or prosocial behavior \cite{schnall2010elevation}. Together, these measures capture both relatively global evaluations and dispositions, such as life satisfaction and trait gratitude, and more immediate emotional responses, such as positive affect, negative affect, and moral elevation.

After writing, participants report their writing self-efficacy using items adapted from Schwarzer and Renner \cite{schwarzer2009health}. These items assess participants' confidence in completing the writing task and overcoming writing-related difficulties. Participants also complete adapted Likert measures of \emph{ownership}, or the extent to which the resulting text feels like their own; \emph{control}, or their perceived influence over the writing process and final text; and \emph{text match}, or how well the final text reflects what they intended to express. Complementing these absolute ratings, Me--AI sliders assess the relative contribution attributed to the participant versus the AI in terms of ownership, control, and leadership \cite{draxler2024ai}.

Furthermore, we assess their experience with the AI agent using the User Experience Questionnaire (UEQ) \cite{laugwitz2008construction}, which captures attractiveness, perspicuity, efficiency, dependability, stimulation, and novelty, and the System Usability Scale (SUS) \cite{lewis2018system}, which provides an overall measure of perceived system usability and ease of use. Full item wording and scoring procedures are provided in \Cref{app:measures}.

\subsection{Behavioral Measures}

\subsubsection{Writing Interaction Behavior}
We derive behavioral engagement measures from the writing-interface logs. For co-writing, we record the number of autocomplete suggestions shown, accepted, and rejected, and calculate the resolved acceptance rate as $\text{accepted}/(\text{accepted}+\text{rejected})$. For scaffolding, we record participants' likes and dislikes of reflective prompts and calculate the like rate as $\text{likes}/(\text{likes}+\text{dislikes})$. The acceptance rate is computed for co-writing participants who receive at least one autocomplete suggestion, whereas the like rate is computed for scaffolding participants who rate at least one prompt. We also record each participant's final submitted word count.

\subsubsection{Self-disclosure Levels}
To characterize the content of participants' writing, we use GPT-5.2-chat deployed through Azure \cite{azuregpt5} to score each final text on three dimensions of self-disclosure (\emph{feelings}, \emph{information}, and \emph{thoughts}) using a three-point rubric adapted from prior self-disclosure coding schemes \cite{tran2025robot}. \emph{Feelings} range from matter-of-fact descriptions without affective content (score 1) to the disclosure of deep emotions (score 3). \emph{Information} ranges from general statements without personal details (score 1) to highly personal revelations about the writer or close others (score 3). \emph{Thoughts} capture the depth of personal ideas and reflections, ranging from general ideas with no personal reference (score 1) to intimate or aspirational reflections on personal characteristics and past or future experiences (score 3). We treat these scores as proxy annotations of disclosure content rather than human-coded evaluations of writing quality. To reduce run-to-run variability, we rate each text five times and use the mean score for each dimension in the analysis.

\subsubsection{Open-ended Response}
Finally, we collect qualitative feedback through four open-ended questions about what they liked (``What did you like about the system?''), disliked (``What did you not like about the system?'') about the system, what could be improved (``Please share anything that can be improved about the system.''),  and any additional thoughts about the writing experience (``Please share any other thoughts or feedback you have about the writing experience.''). We manually reviewed the responses to contextualize the quantitative findings and identify illustrative accounts of participants' experiences. Full descriptions of all measures are provided in \Cref{app:measures}.

\section{Results}
In our analysis, we first examined changes in well-being and whether these changes differed by writing format and AI assistance mode (RQ1), followed by writing experience (RQ2), and self-disclosure in the final texts (RQ3). We then report exploratory associations between participants' interactions with AI and their experiences. We did not detect an interaction between writing format and AI-assistance mode in pre--post well-being changes, although each factor was associated with differences in particular outcomes.

\subsection{RQ1: Changes in Well-Being Based on Writing Types and AI-Support Modes}

\begin{figure*}[t]
\centering
\includegraphics[width=\textwidth]{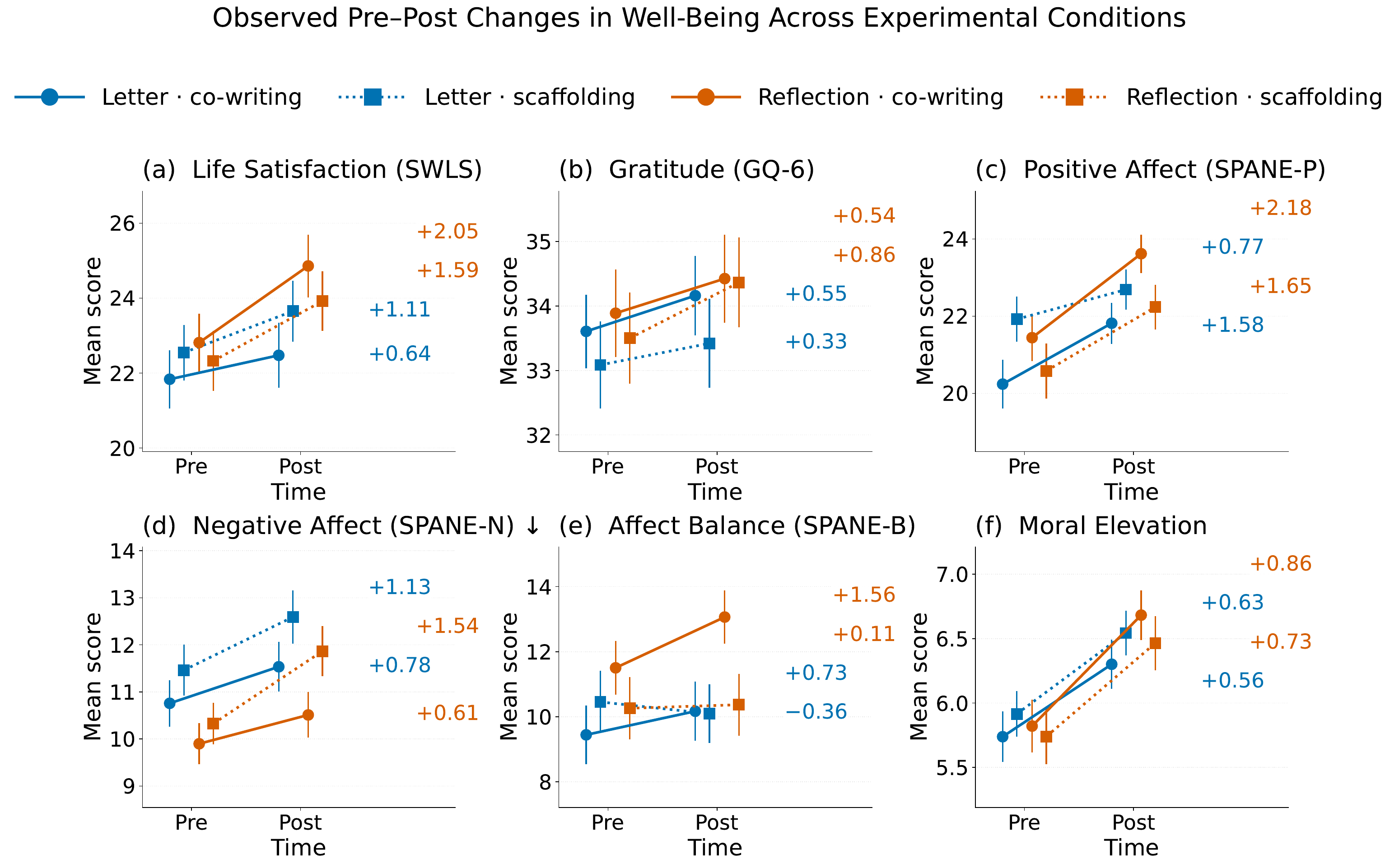}
\caption{Changes in pre--post well-being scores by gratitude-writing format and AI-assistance mode. Panels denote (a) life satisfaction (SWLS), (b) trait gratitude (GQ-6), (c) positive affect (SPANE-P), (d) negative affect (SPANE-N), (e) affect balance (SPANE-B), and (f) moral elevation. Points represent observed condition means; error bars indicate SE. For SPANE-N, higher scores indicate greater negative affect.  Overall, life satisfaction, gratitude, positive affect, and moral elevation increase following the intervention; however, negative affect also increases, resulting in only a small change in affect balance.}
\Description{Six line charts arranged in two rows show pre-test and post-test condition means. Panel a presents life satisfaction, panel b trait gratitude, panel c positive affect, panel d negative affect, panel e affect balance, and panel f moral elevation. Blue lines represent gratitude-letter writing, and orange lines represent reflection. Solid lines with circular markers represent co-writing, while dotted lines with square markers represent scaffolding. Vertical error bars surround each mean, and numerical labels report the pre-to-post change. Most lines rise in panels a through d and f; panel e shows smaller, mixed changes across conditions.}
\label{fig:wellbeingchangeoverall}
\end{figure*}

To evaluate whether pre--post changes vary across conditions, we fitted a separate $2 \times 2 \times 2$ linear mixed-effects model (LMM) for each outcome. Each model included Time (\textit{pre-test} vs.\ \textit{post-test}), writing type (\textit{gratitude letter} vs.\ \textit{gratitude reflection}), and AI-support (\textit{co-writing} vs.\ \textit{scaffolding}) as fixed effects, with a random intercept for each participant. We used Wald tests to assess each main effect and interaction. Because all factors had two levels, each test had one degree of freedom and is reported as $\chi^2(1)$. Full results are reported in \Cref{tab:rq1}, and observed condition means are shown in \Cref{fig:wellbeingchangeoverall}.%Results are reported in 

\subsubsection{Satisfaction With Life (SWLS)}
We found a significant main effect of time, $\chi^2(1)=57.75$, $p<0.001$, and a significant time $\times$ writing type interaction, $\chi^2(1)=5.78$, $p=0.016$. Model-implied estimates indicated larger SWLS increases in the reflection conditions ($\Delta=1.65$--$2.17$) than in the gratitude-letter conditions ($\Delta=0.78$--$1.21$). Neither the main effect of writing type nor that of AI-support was significant. The time $\times$ AI-support interaction, writing type $\times$ AI-support interaction, and three-way interaction were also nonsignificant. 

\subsubsection{Trait Gratitude (GQ-6).}
Trait gratitude increased similarly across all conditions. Specifically, only the main effect of time was significant, $\chi^2(1)=10.74$, $p=0.001$. Neither writing type nor AI-support had a significant main effect. None of the interactions was significant: time $\times$ writing type, time $\times$ AI-support, writing type $\times$ AI-support, or the three-way interaction.

\subsubsection{Affect Balance (SPANE-B)}

For affect balance, we found a significant time $\times$ AI-support interaction, $\chi^2(1)=7.19$, $p=0.007$. Affect balance improved with co-writing ($\Delta=0.83$--$1.55$) but not with scaffolding ($\Delta=-0.46$--$0.00$). The main effects of time, writing type, and AI-support were nonsignificant. The remaining interactions are also nonsignificant: time $\times$ writing type, writing type $\times$ AI-support, and the three-way interaction.

To clarify this pattern, we examined positive and negative affect separately. For positive affect (SPANE-P), we found a significant main effect of time, $\chi^2(1)=81.48$, $p<0.001$, a significant time $\times$ AI-support interaction, $\chi^2(1)=4.33$, $p=0.037$, and a significant writing type $\times$ AI-support interaction, $\chi^2(1)=4.37$, $p=0.036$. Neither the main effect of writing type nor that of AI-support was significant. The time $\times$ writing type interaction, and the three-way interaction were also not significant. For negative affect (SPANE-N), we found significant main effects of time, $\chi^2(1)=35.11$, $p<0.001$, and writing type, $\chi^2(1)=4.30$, $p=0.038$. The main effect of AI-support approaches did not reach statistical significance. None of the interactions was significant: time $\times$ writing type, time $\times$ AI-support, writing type $\times$ AI-support, or the three-way interaction.

\subsubsection{Moral Elevation.}
Only the main effect of Time was significant, $\chi^2(1)=149.96$, $p<0.001$. Neither writing type nor AI-support had a significant main effect. All interactions were also nonsignificant: Time $\times$ writing type, Time $\times$ AI-support, writing type $\times$ AI-support, and the three-way interaction.

\subsubsection{Summary of Results}
Changes after gratitude writing showed that gratitude reflection led to larger gains in life satisfaction (SWLS) than gratitude letter writing, while co-writing led to larger gains in positive feelings (SPANE-P) and the balance of positive over negative feelings (SPANE-B) than scaffolding. Participants also reported greater gratitude (GQ-6) and moral elevation, but more negative feelings (SPANE-N), without clear differences in these changes between conditions. We found no evidence that the benefits of either AI-assistance mode depended on whether participants wrote a reflection or a letter.

\begin{table*}[t]
\centering
\caption{Omnibus Wald tests from the $2\times2\times2$ well-being LMMs, reported as $\chi^2(1)$ ($p$). T~=~Time, W~=~Gratitude Writing Type, S~=~AI-Support Mode (RQ1). Bold marks $p<0.05$ ($^{*}p<0.05$, $^{**}p<0.01$, $^{***}p<0.001$); all tests are unadjusted. Most well-being outcomes rise over time; reflection yields larger SWLS gains, co-writing yields larger SPANE-P and SPANE-B gains, and GQ-6 and moral elevation increase similarly across conditions.}
\label{tab:rq1}
\small
\setlength{\tabcolsep}{4pt}
\renewcommand{\arraystretch}{1.08}
\resizebox{\textwidth}{!}{%
\begin{tabular}{@{}lccccccc@{}}
\toprule
Outcome & T & W & S & T$\times$W & T$\times$S & W$\times$S & T$\times$W$\times$S \\
\midrule
Life Satisfaction (SWLS)
  & $\mathbf{57.75^{***}\;(<0.001)}$
  & $1.36\;(0.244)$
  & $0.00\;(0.968)$
  & $\mathbf{5.78^{*}\;(0.016)}$
  & $0.01\;(0.914)$
  & $0.93\;(0.336)$
  & $1.55\;(0.213)$ \\
Trait Gratitude (GQ-6)
  & $\mathbf{10.74^{**}\;(0.001)}$
  & $0.51\;(0.473)$
  & $0.38\;(0.539)$
  & $0.17\;(0.683)$
  & $0.23\;(0.629)$
  & $0.08\;(0.779)$
  & $0.21\;(0.648)$ \\
Positive Affect (SPANE-P)
  & $\mathbf{81.48^{***}\;(<0.001)}$
  & $0.18\;(0.672)$
  & $0.01\;(0.904)$
  & $2.87\;(0.090)$
  & $\mathbf{4.33^{*}\;(0.037)}$
  & $\mathbf{4.37^{*}\;(0.036)}$
  & $0.37\;(0.542)$ \\
Negative Affect (SPANE-N)
  & $\mathbf{35.11^{***}\;(<0.001)}$
  & $\mathbf{4.30^{*}\;(0.038)}$
  & $3.50\;(0.061)$
  & $0.01\;(0.907)$
  & $2.97\;(0.085)$
  & $0.00\;(0.967)$
  & $0.70\;(0.401)$ \\
Affect Balance (SPANE-B)
  & $3.26\;(0.071)$
  & $2.03\;(0.154)$
  & $0.80\;(0.371)$
  & $1.26\;(0.262)$
  & $\mathbf{7.19^{**}\;(0.007)}$
  & $1.88\;(0.170)$
  & $0.06\;(0.805)$ \\
Moral Elevation
  & $\mathbf{149.96^{***}\;(<0.001)}$
  & $0.06\;(0.809)$
  & $0.02\;(0.894)$
  & $2.60\;(0.107)$
  & $0.13\;(0.722)$
  & $0.86\;(0.353)$
  & $0.42\;(0.518)$ \\
\bottomrule
\end{tabular}}
\end{table*}

%\subsection{Writing Experience} %\cite{schwarzer2009health}
\subsection{RQ2: Writing Experience and Perceived Agency} 
To examine how the experimental conditions shape participants' writing experiences, we conducted separate HC3-robust Type III writing type $\times$ AI-support ANOVAs for writing self-efficacy, system perceptions, ownership, control, and human-agency attribution. Tukey's HSD tests were used for post-hoc pairwise comparisons following significant ANOVA tests. The ANOVA results are shown in \Cref{tab:rq2}.

\subsubsection{User Experience and Usability}

For the User Experience Questionnaire (UEQ), we found a significant main effect of AI-support type, $F(1,371)=9.25$, $p=0.003$, $\eta_p^2=0.024$. Neither the main effect of writing type nor the interaction was significant. Tukey HSD comparisons showed that the gratitude reflection/co-writing condition scored higher than both the gratitude-letter/scaffolding condition ($\Delta=0.45$, $p_{\mathrm{adj}}=0.030$) and the gratitude reflection/scaffolding condition ($\Delta=0.51$, $p_{\mathrm{adj}}=0.014$).
For the System Usability Scale (SUS), we found a significant main effect of AI-support type, $F(1,375)=8.69$, $p=0.003$, $\eta_p^2=0.023$. Neither the main effect of writing type nor the interaction was significant. Tukey HSD comparisons indicated that the gratitude reflection/co-writing condition scored significantly higher than the gratitude reflection/scaffolding condition ($\Delta=7.04$, $p_{\mathrm{adj}}=0.044$). All other pairwise SUS comparisons were nonsignificant.

%Overall, \textbf{gratitude-letter writers report greater perceived control than reflection writers, driven by lower control in the gratitude reflection/co-writing condition relative to both gratitude-letter conditions; Likert ownership shows no significant condition effects.}

\subsubsection{Ownership, Control, and Leadership}

\paragraph{Ownership and Control}

For perceived control, we found a significant main effect of writing type, $F(1,376)=6.25$, $p=0.013$, $\eta_p^2=0.016$. Neither the main effect of AI-support type nor the interaction was significant. Tukey HSD comparisons showed that the gratitude reflection/co-writing condition scored lower than both the gratitude-letter/co-writing condition ($\Delta=-0.32$, $p_{\mathrm{adj}}=0.030$) and the gratitude-letter/scaffolding condition ($\Delta=-0.30$, $p_{\mathrm{adj}}=0.047$).
For perceived ownership, we found no significant main effect of writing type or AI-support type. The interaction was also nonsignificant.

\begin{figure*}[t]

\centering

\includegraphics[width=\textwidth]{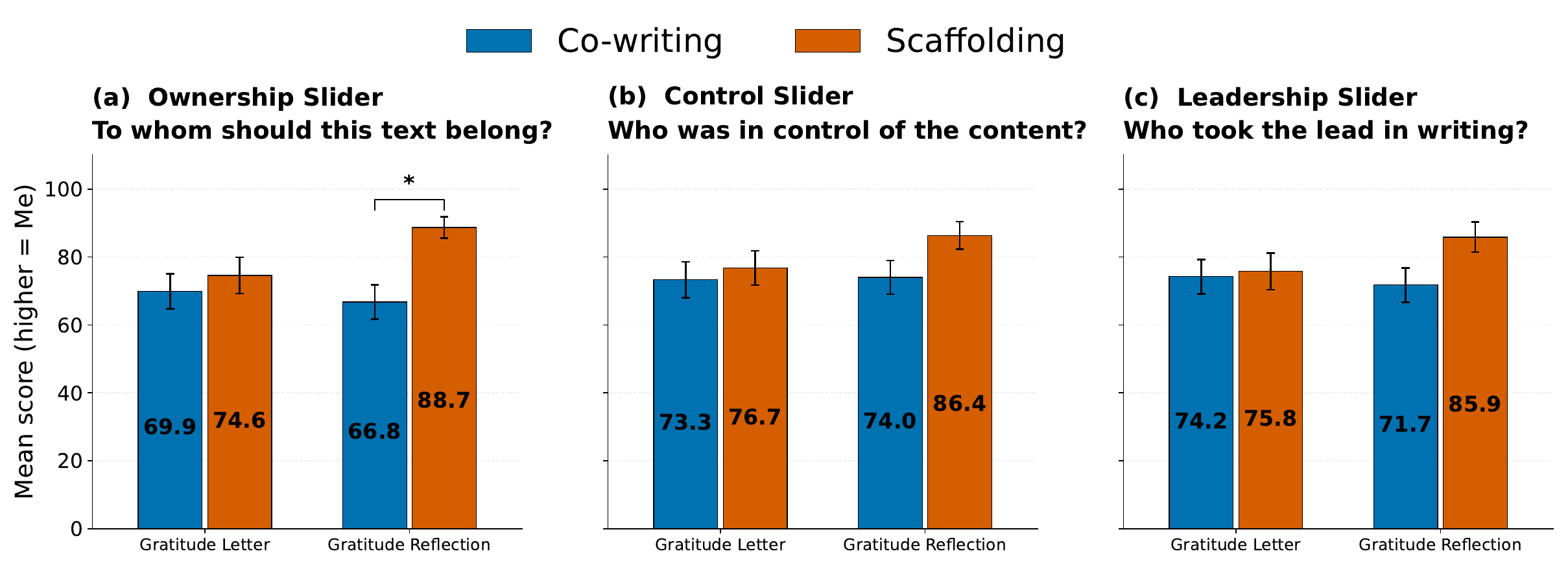}

\caption{Ownership, Control, and Leadership Slider ratings by writing task and AI-support type. Bars show condition means ($\pm$ SE) for Gratitude Letter and Gratitude Reflection under Co-writing (blue) and Scaffolding (orange). Higher scores indicate greater attribution to the participant (Me). The only significant within-task Tukey contrast: Ownership under Gratitude Reflection, where scaffolding exceeds co-writing ($*=p_{\mathrm{adj}}<.05$). Nonsignificant comparisons are omitted.}

\Description{Three grouped bar charts for the Ownership Slider, Control Slider, and Leadership Slider. The x-axis compares Gratitude Letter versus Gratitude Reflection; blue bars are co-writing and orange bars are scaffolding. Means are printed inside bars with SE caps. A single asterisk bracket marks the significant Ownership contrast under Gratitude Reflection (scaffolding higher than co-writing); nonsignificant comparisons are not annotated.}
\label{fig:meaisliders}

\end{figure*}

\paragraph{Relative Human--AI Attribution}
Following \cite{draxler2024ai}, we analyzed the ownership, control, and leadership sliders separately. For the inferential analysis of human ownership, we oriented the scores so that higher values indicated greater attribution to the participant. \Cref{fig:meaisliders} showed condition means with standard error bars for the ownership, control, and leadership sliders; higher scores indicate greater attribution to the participant. 
For the ownership slider, there was a significant main effect of AI-support type, $F(1,379)=7.74$, $p=0.006$, $\eta_p^2=.020$. Neither the main effect of writing type nor the interaction was significant. Tukey HSD post-hoc comparisons revealed that the gratitude reflection/scaffolding condition scored significantly higher than both the gratitude-letter/co-writing condition ($\Delta=18.74$, $p_{\mathrm{adj}}=0.035$) and the gratitude reflection/co-writing condition ($\Delta=21.92$, $p_{\mathrm{adj}}=0.011$). All other pairwise comparisons were nonsignificant.
For the control slider, neither writing type, AI-support type nor their interaction was significant.
For the leadership slider, we found no significant main effect of writing type, or AI-support type, and no significant interaction.

\subsubsection{Writing Self-Efficacy and Text Match}

\paragraph{Text Match}
For perceived text match score, there was a significant main effect of writing type, $F(1, 379) = 5.40$, $p = .021$, $\eta_p^2 = .014$. The main effect of AI-support Type and the interaction were not significant. Despite the significant omnibus main effect of writing type, subsequent Tukey HSD comparisons did not reveal significant differences between any specific condition pairs (all $p_{\mathrm{adj}}\geq.081$).

\paragraph{Writing Self-Efficacy}
For total writing self-efficacy, neither the main effect of writing type nor that of AI-support type was significant. Their interaction was also nonsignificant. Writing self-efficacy had two components: writing-process efficacy and writing-barrier efficacy. For writing-process efficacy, we found no significant main effect of writing type or AI-support and no significant interaction. For writing-barrier efficacy, neither the writing type nor AI-support type had a significant main effect. Their interaction was also nonsignificant. 

\begin{table*}[t]
\centering
\caption{HC3-robust Type~III ANOVAs for writing-experience outcomes (RQ2). Cells report $F(1,\mathit{df})$ ($p$) [$\eta_p^2$]; $\mathit{df}$ is the residual degrees of freedom. Bold marks $p<0.05$ ($^{*}p<0.05$, $^{**}p<0.01$, $^{***}p<0.001$). Direction notes significant marginal main effects (not every pairwise contrast). Co-writing improves UEQ and SUS, scaffolding increases human ownership on the Me--AI slider, and letter writing yields higher Likert control and text-match ratings, while writing self-efficacy and the other agency sliders do not differ by condition.}
\label{tab:rq2}
\small
\setlength{\tabcolsep}{3.5pt}
\renewcommand{\arraystretch}{1.08}
\resizebox{\textwidth}{!}{%
\begin{tabular}{@{}lccccc@{}}
\toprule
Outcome & $\mathit{df}$ & Gratitude Writing Type & AI-support Mode & W$\times$S & Direction \\
\midrule
User Experience Questionnaire (UEQ)
  & 371
  & $0.67\;(0.413)\;[0.002]$
  & $\mathbf{9.25^{**}\;(0.003)\;[0.024]}$
  & $1.70\;(0.193)\;[0.005]$
  & Co-writing $>$ scaffolding \\
System Usability Scale (SUS)
  & 375
  & $0.18\;(0.672)\;[<0.001]$
  & $\mathbf{8.69^{**}\;(0.003)\;[0.023]}$
  & $0.66\;(0.416)\;[0.002]$
  & Co-writing $>$ scaffolding \\
Text-match Score 
  & 379
  & $\mathbf{5.40^{*}\;(0.021)\;[0.014]}$
  & $0.18\;(0.669)\;[<0.001]$
  & $1.10\;(0.294)\;[0.003]$
  & Letter $>$ reflection \\
Control Score
  & 376
  & $\mathbf{6.25^{*}\;(0.013)\;[0.016]}$
  & $1.27\;(0.261)\;[0.003]$
  & $1.87\;(0.172)\;[0.005]$
  & Letter $>$ reflection \\
Ownership Score
  & 376
  & $0.25\;(0.619)\;[0.001]$
  & $0.94\;(0.333)\;[0.002]$
  & $0.52\;(0.473)\;[0.001]$
  & No significant terms \\
Ownership Slider
  & 379
  & $1.31\;(0.253)\;[0.003]$
  & $\mathbf{7.74^{**}\;(0.006)\;[0.020]}$
  & $3.28\;(0.071)\;[0.009]$
  & Scaffolding $>$ co-writing \\
Control Slider
  & 379
  & $1.12\;(0.291)\;[0.003]$
  & $2.64\;(0.105)\;[0.007]$
  & $0.84\;(0.361)\;[0.002]$
  & No significant terms \\
Leadership Slider
  & 379
  & $0.57\;(0.450)\;[0.002]$
  & $2.44\;(0.119)\;[0.006]$
  & $1.57\;(0.211)\;[0.004]$
  & No significant terms \\
Writing Efficacy: total
  & 395
  & $2.75\;(0.098)\;[0.007]$
  & $1.56\;(0.213)\;[0.004]$
  & $0.19\;(0.662)\;[0.000]$
  & No significant terms \\
Writing Efficacy: process
  & 395
  & $2.14\;(0.145)\;[0.005]$
  & $1.30\;(0.255)\;[0.003]$
  & $1.25\;(0.263)\;[0.003]$
  & No significant terms \\
Writing Efficacy: barriers
  & 396
  & $2.38\;(0.124)\;[0.006]$
  & $1.21\;(0.273)\;[0.003]$
  & $2.28\;(0.132)\;[0.006]$
  & No significant terms \\
\bottomrule
\end{tabular}}
\end{table*}

\subsubsection{Summary of Results}
Our analysis showed that participants rated co-writing more positively for user experience (UEQ) and usability (SUS), while those using scaffolding attributed more ownership of the text to themselves rather than the AI. This ownership difference appeared on the Me-AI slider but not on the separate ownership scale. Letter writers reported greater perceived control and higher text-match ratings than reflection writers. We found no significant difference in writing self-efficacy or in the relative attribution of control or leadership to the participant versus AI. No significant interactions indicated that the effects of AI-assistance mode depended on the writing format.

\subsection{RQ3: Disclosures in the Final Writing}

Similar to \cite{tran2025robot}, we used an LLM to evaluate self-disclosure in participants' final texts. GPT-5.2-chat was used to rate each text five times on three self-disclosure dimensions: feelings, information, and thoughts. To evaluate the consistency across repeated model evaluation, we calculated the proportion of ratings that matched each text's most frequent score. Mean agreement was 96.9\% for feelings, 98.9\% for information, and 93.1\% for thoughts. For 255 of 401 texts ($63.6\%$), all five ratings were identical within each of the three dimensions. \Cref{tab:disclosure-descriptives} reported the condition-level descriptives and marginal summaries by writing type and AI-support type. We then conducted separate HC3-robust Type III two-way ANOVAs with writing type and AI-support mode for each dimension, using the mean of the five ratings, and for overall disclosure, averaged across dimensions. Tukey HSD comparisons followed significant omnibus effects; full ANOVA results are reported in \Cref{tab:rq3}.  

\begin{table*}[t]
\centering
\caption{LLM-rated self-disclosure by experimental condition and marginal factor level. Values are mean $\pm$ SD on a 1--3 scale; participant-level scores are the means across five scoring passes. \protect\textcolor{blue}{Blue} indicates significantly higher feelings disclosure for gratitude-letters, and \protect\textcolor{orange}{orange} indicates significantly higher thoughts disclosure for gratitude reflections. Gratitude reflection has higher scores in thoughts disclosure, while gratitude-letter has higher scores in feelings disclosure; factual information disclosure scores do not differ significantly across all conditions.}
\label{tab:disclosure-descriptives}
%\small
\setlength{\tabcolsep}{3pt}
\renewcommand{\arraystretch}{1.05}
\begin{tabular}{@{}lccc@{}}
\toprule
Condition & Feelings & Thoughts & Information \\
\midrule
\multicolumn{4}{@{}l}{\textit{Experimental Conditions}} \\
gratitude-letter / co-writing ($n=104$)
& \textcolor{blue}{$2.87 \pm 0.35$} & \textcolor{orange}{$2.45 \pm 0.46$} & $2.94 \pm 0.26$ \\
gratitude-letter / scaffolding ($n=107$)
& \textcolor{blue}{$2.82 \pm 0.39$} & \textcolor{orange}{$2.50 \pm 0.48$} & $2.95 \pm 0.25$ \\
gratitude reflection / co-writing ($n=98$)
& \textcolor{blue}{$2.67 \pm 0.43$} & \textcolor{orange}{$2.73 \pm 0.39$} & $2.90 \pm 0.28$ \\
gratitude reflection / scaffolding ($n=92$)
& \textcolor{blue}{$2.63 \pm 0.49$} & \textcolor{orange}{$2.70 \pm 0.44$} & $2.89 \pm 0.33$ \\
\addlinespace[4pt]
\multicolumn{4}{@{}l}{\textit{AI-Support Type}} \\
co-writing ($n=202$)
& $2.77 \pm 0.40$ & $2.59 \pm 0.45$ & $2.92 \pm 0.27$ \\
scaffolding ($n=199$)
& $2.73 \pm 0.45$ & $2.59 \pm 0.48$ & $2.92 \pm 0.29$ \\
\addlinespace[4pt]
\multicolumn{4}{@{}l}{\textit{Gratitude Writing Type}} \\
gratitude-letter ($n=211$)
& \textcolor{blue}{$2.84 \pm 0.37$} & \textcolor{orange}{$2.47 \pm 0.47$} & $2.94 \pm 0.26$ \\
gratitude reflection ($n=190$)
& \textcolor{blue}{$2.65 \pm 0.46$} & \textcolor{orange}{$2.72 \pm 0.42$} & $2.90 \pm 0.31$ \\
\bottomrule
\end{tabular}
\end{table*}

\begin{table*}[t]
\centering
\caption{HC3-robust Type~III ANOVAs for LLM-rated self-disclosure, reported as $F(1,397)$ ($p$). Bold marks $p<0.05$ ($^{*}p<0.05$, $^{**}p<0.01$, $^{***}p<0.001$); all tests are unadjusted. Outcomes are means across five scoring passes. Writing type shapes what participants disclose, letters disclose more feelings and reflections more thoughts, whereas AI-support type does not affect disclosure scores.}
\label{tab:rq3}
\small
\setlength{\tabcolsep}{5pt}
\renewcommand{\arraystretch}{1.08}
\begin{tabular}{@{}lcccc@{}}
\toprule
Outcome & Writing type & AI-support & W$\times$S & Summary \\
\midrule
Feelings
  & $\mathbf{21.60^{***}\;(<0.001)}$
  & $0.94\;(0.334)$
  & $0.02\;(0.888)$
  & Letter $>$ reflection \\
Thoughts
  & $\mathbf{30.22^{***}\;(<0.001)}$
  & $0.04\;(0.834)$
  & $0.80\;(0.372)$
  & Reflection $>$ letter \\
Information
  & $2.61\;(0.107)$
  & $0.01\;(0.916)$
  & $0.23\;(0.632)$
  & No significant terms \\
Overall disclosure
  & $<0.01\;(0.979)$
  & $0.10\;(0.752)$
  & $0.28\;(0.597)$
  & No significant terms \\
\bottomrule
\end{tabular}
\end{table*}

\subsubsection{Feelings}

For feelings disclosure, we found a significant main effect of writing type, $F(1,397)=21.60$, $p<0.001$. Neither the main effect of AI-support type nor the Writing Type $\times$ AI-Support Type interaction was significant. Tukey HSD comparisons showed that gratitude-letter writers disclosed more feelings than gratitude-reflection writers in all four cross-writing-type comparisons (all $p_{\mathrm{adj}}<0.05$). 

\subsubsection{Thoughts}

For thoughts disclosure, we found a significant main effect of writing type in the opposite direction, $F(1,397)=30.22$, $p<.001$. Neither the main effect of AI-support type nor the Writing Type $\times$ AI-Support Type interaction was significant. Tukey HSD comparisons showed that gratitude-reflection writers disclosed more thoughts than gratitude-letter writers across both AI-support types (all cross-writing-type $p_{\mathrm{adj}}<.05$). 

\subsubsection{Information}

For factual information disclosure, neither the main effect of writing type nor that of AI-support type was significant. The Writing Type $\times$ AI-Support Type interaction was also nonsignificant. 

\subsubsection{Overall Disclosure Level} 
Similarly, for overall disclosure, calculated as the mean of feelings, information, and thoughts, neither the main effect of writing type, $F(1,397)<0.01$, $p=.979$, nor that of AI-support type, $F(1,397)=0.10$, $p=.752$, was significant. The Writing Type $\times$ AI-Support Type interaction was also nonsignificant, $F(1,397)=0.28$, $p=.597$.

\subsubsection{Summary of Results}
Our analysis of LLM-rated self-disclosure showed that gratitude letters expressed deeper feelings, while gratitude reflections contained deeper disclosure of personal thoughts. Neither writing format nor AI-assistance mode significantly affected factual information disclosure or the overall disclosure score. No significant AI-assistance effects or interactions were found.

\subsection{Exploratory Analysis: AI Suggestion Acceptance and Prompt Ratings During Gratitude Writing}

We conducted exploratory analyses to examine how participants' suggestion acceptance in \textit{co-writing} condition and prompt ratings in \textit{scaffolding} condition are related to their well-being changes, writing experience, and perceived agency. These measures capture different responses to AI assistance: acceptance indicates incorporation of suggested text, whereas prompt ratings indicate perceived helpfulness. For each AI-assistance mode, Spearman correlation coefficients were calculated to examine how suggestion acceptance or prompt ratings are associated with changes in well-being and post-writing experience scores. Benjamini-Hochberg corrections were applied to account for multiple comparisons. 

\subsubsection{Suggestion Acceptance in Co-Writing}

We tested whether autocomplete acceptance differed between gratitude-letter and reflection writing and how it related to writing experience, perceived agency, and changes in well-being. We defined the resolved acceptance rate as the number of accepted suggestions divided by the total number of suggestions accepted or rejected. Because autocomplete was available only in co-writing, these analyses were restricted to co-writing participants. Among the 199 participants who received at least one suggestion, \textbf{gratitude-reflection writers show descriptively higher acceptance rates ($M=0.18$, $n=97$) than gratitude-letter writers ($M=0.12$, $n=102$).}

We pooled participants across the two co-writing conditions and computed Spearman correlations between the resolved acceptance rate, post-task writing experience, perceived agency, and pre--post well-being changes. We applied Benjamini--Hochberg false-discovery-rate correction within this family of comparisons. \Cref{tab:cowriting-correlations} reported the results.

\definecolor{darkgreen}{HTML}{006400}
\definecolor{darkred}{HTML}{8B0000}

\begin{table*}[t]
\centering
\caption{Spearman associations between co-writing acceptance rate, writing experience, and well-being outcomes. \protect\textcolor{darkgreen}{Green} indicates a significant positive association and \protect\textcolor{darkred}{red} indicates a significant negative association after Benjamini-Hochberg FDR correction ($p_{\mathrm{FDR}}<.05$); black indicates a nonsignificant association. A higher acceptance rate is associated with higher UEQ ratings but lower perceived ownership, control, and human leadership.}
\label{tab:cowriting-correlations}
\setlength{\tabcolsep}{4pt}
\renewcommand{\arraystretch}{1.05}
\begin{tabular}{lrr}
\toprule
Outcome & $\rho$ & $p_{\mathrm{FDR}}$ \\
\midrule
Ownership Slider
& \textcolor{darkred}{$-0.429$}
& \textcolor{darkred}{$<0.001$} \\
Control Slider
& \textcolor{darkred}{$-0.243$}
& \textcolor{darkred}{$0.004$} \\
Leadership Slider
& \textcolor{darkred}{$-0.349$}
& \textcolor{darkred}{$<0.001$} \\
Ownership Score
& \textcolor{darkred}{$-0.239$}
& \textcolor{darkred}{$0.004$} \\
Control Score
& \textcolor{darkred}{$-0.292$}
& \textcolor{darkred}{$<0.001$} \\
User Experience Questionnaire (UEQ)
& \textcolor{darkgreen}{$0.221$}
& \textcolor{darkgreen}{$0.009$} \\
System Usability Scale (SUS)
& $0.119$ & $0.208$ \\
Total Writing self-efficacy score
& $-0.153$ & $0.094$ \\
Text-match Score 
& $-0.068$ & $0.554$ \\
Change in Life satisfaction (SWLS)
& $0.140$ & $0.128$ \\
Change in Affect balance (SPANE-B)
& $0.049$ & $0.683$ \\
Change in Trait gratitude (GQ-6)
& $0.034$ & $0.758$ \\
Change in Moral elevation
& $0.032$ & $0.768$ \\
\bottomrule
\end{tabular}
\end{table*}

Higher acceptance rates were associated with weaker perceived human ownership, control, and leadership across the three sliders and the ownership and control composites ($\rho=-0.429$ to $-0.239$, all $p_{\mathrm{FDR}}\leq 0.004$). The strongest association was with the ownership slider ($\rho=-0.429$, $p_{\mathrm{FDR}}<0.001$).

Moreover, higher acceptance rates were also associated with more positive user experience, as measured by UEQ ($\rho=0.221$, $p_{\mathrm{FDR}}=0.009$), but were not significantly associated with usability, writing self-efficacy, or text match after correction (all $p_{\mathrm{FDR}}\geq 0.094$). Associations with changes in life satisfaction, affect balance, trait gratitude, and moral elevation were likewise nonsignificant (all $p_{\mathrm{FDR}}\geq 0.128$).

Together, these findings link \textbf{greater suggestion acceptance to more positive user experience and weaker perceived human agency but provide no evidence that greater acceptance is associated with larger short-term well-being improvements}. These associations do not establish causal direction.

\subsubsection{Prompt Helpfulness Ratings in Scaffolding}

We examined whether participants' evaluations of scaffolding prompts differed between gratitude-letter and reflection writing and how these evaluations related to writing experience, perceived agency, and well-being changes. We defined the like rate as the number of likes divided by the total number of likes and dislikes.

Among the 110 scaffolding participants who rated at least one prompt, \textbf{gratitude-reflection writers ($M=0.87$, $n=54$) and gratitude-letter writers ($M=0.86$, $n=56$) show descriptively similar like rates.}

We pooled participants across the two scaffolding conditions and computed Spearman correlations between the like rate, post-task writing experience, perceived agency, and pre--post well-being changes. These analyses included only participants who rated at least one prompt. We applied Benjamini--Hochberg false-discovery-rate correction within this family of comparisons. \Cref{tab:scaffolding-correlations} reported the results.

\begin{table*}[t]
\centering
\caption{Spearman associations between scaffolding like rate, writing experience, and well-being outcomes. \protect\textcolor{darkgreen}{Green} indicates a significant positive association; 
black indicates a nonsignificant association. A higher like rate is associated with higher UEQ ratings, but shows no significant relationship with perceived agency or well-being.}
\label{tab:scaffolding-correlations}
\setlength{\tabcolsep}{4pt}
\renewcommand{\arraystretch}{1.05}
\begin{tabular}{lrr}
\toprule
Outcome & $\rho$ & $p_{\mathrm{FDR}}$ \\
\midrule
Ownership slider
& $-0.041$
& $0.768$ \\
Control slider
& $-0.010$
& $0.935$ \\
Leadership slider
& $-0.058$
& $0.736$ \\
Ownership score
& $-0.110$
& $0.539$ \\
Control score
& $-0.046$
& $0.761$ \\
User Experience Questionnaire (UEQ)
& \textcolor{darkgreen}{$0.331$}
& \textcolor{darkgreen}{$0.021$} \\
System Usability Scale (SUS)
& $0.094$ & $0.602$ \\
Total Writing self-efficacy score
& $0.174$ & $0.249$ \\
Text-match score
& $0.060$ & $0.736$ \\
Change in Life satisfaction (SWLS)
& $0.071$ & $0.717$ \\
Change in Affect balance (SPANE-B)
& $0.146$ & $0.389$ \\
Change in Trait gratitude (GQ-6)
& $-0.044$ & $0.761$ \\
Change in Moral elevation
& $-0.089$ & $0.627$ \\
\bottomrule
\end{tabular}
\end{table*}

The like rate was not significantly associated with perceived human ownership, control, or leadership across the three sliders or the ownership and control composites after correction ($\rho=-0.110$ to $-0.010$, all $p_{\mathrm{FDR}}\geq 0.539$).

Moreover, higher like rates were associated with more positive user experience, as measured by UEQ ($\rho=0.331$, $p_{\mathrm{FDR}}=0.021$). Associations with usability, writing self-efficacy, and text match were not significant after correction (all $p_{\mathrm{FDR}}\geq 0.249$).

As with the co-writing acceptance rate, the scaffolding like rate was not significantly associated with changes in life satisfaction, affect balance, trait gratitude, or moral elevation after correction (all $p_{\mathrm{FDR}}\geq 0.389$).

Together, \textbf{these findings link more favorable evaluations of scaffolding prompts to more positive user experience, without significant associations with perceived agency or short-term well-being changes.} These results concern evaluations among participants who rate prompts rather than how frequently participants use or act on the guidance.

\subsection{Qualitative Feedback}
\label{sec:qual-results}
We analyzed $1,241$ responses to the four open-ended questions using LLM-assisted multi-label coding (Claude Sonnet 5, \cite{anthropic2026claudesonnet5}) with a predefined set of 36 themes. Each response was coded in three runs with shuffled response order, retaining a theme assignment only when it appeared in at least two runs. Theme frequencies were calculated at the participant level: each participant was counted once per theme if it appeared in any of their responses, with percentages based on the $378$ participants who answered at least one question.

\subsubsection{Shared Benefits: Ideas When Stuck, and Staying in Control}
Participants valued both AI-support models for the same core reason: help developing ideas when they were stuck. Ideational support was the most frequently mentioned benefit (35\% of participants; 40\% in co-writing, 29\% in scaffolding). P00323 (reflection/scaffolding) described the questions as ``a little bump to get going on the writing again when running out of ideas,'' and P00261 (reflection/co-writing) valued that the system ``was able to suggest the sentence continuation when I felt stuck.'' Notably, this support did not require accepting the AI's language. P00170 (reflection/co-writing) explained, ``Most of the time, I didn't take the suggested text, but it spurred my thought process, and helped me stay on track, and kept me from getting stuck.'' 

One in five participants, in roughly equal proportions across the two modes (20\% vs.\ 19\%), spontaneously emphasized that the assistance was optional and left them in control: ``I like that it suggests ideas without adding them in. When I stopped writing, it will show me an idea to continue on but it does not push it on me. I still get the power to chose how I want to continue'' (P00390, letter/co-writing); ``It was just there in the background and only popped up when I was at a point where it seemed like I may be stuck.'' (P00064, letter/scaffolding). Ease of use was mentioned by 27\% of participants and did not differ by mode. 

\subsubsection{Mode-Specific Benefits: Speed and Fit vs.\ Deeper Reflection}
Beyond this shared role, the two modes were praised for different qualities. Co-writing participants more often highlighted the speed of the assistance (13\% vs.\ 4\%) and its fit with their draft (24\% vs.\ 13\%): ``the ai seemed to follow the emotion of my text to generate applicable suggestions to flow with my writing'' (P00254, letter/co-writing). 

Scaffolding participants were six times as likely to say that the assistance made them think more deeply (12\% vs.\ 2\%): ``It was able to understand what i was writing about and ask thought provoking questions that could help me think in my writing task.'' (P00338, letter/scaffolding); ``I like how it took the time to wait for my inputs before making any suggestions. The suggestions it made encouraged me to think deeper.'' (P00448, reflection/scaffolding). P00409 (letter/scaffolding) connected this to ownership: ``I felt like it enhanced my writing abilities and served as an aid but I was fully in control of the process.''

\subsubsection{Mode-Specific Limitations: Intrusive Output vs.\ Absent Support}
The most common complaint overall was generic, repetitive, or off-target output (22\%), reported more often in co-writing (27\% vs.\ 17\%). In co-writing, this usually meant completions that lost the thread of the draft: ``It didn't keep up with the context of what I was writing. It just kept throwing in random sentences, and in the end, sentence fragments.'' (P00083, letter/co-writing). In scaffolding, it meant repetitive questions: ``The tips sometimes were just reworded instead of making new ones for the user'' (P00043, reflection/scaffolding). Co-writing participants were also more likely to report that suggestions interrupted their train of thought (13\% vs.\ 6\%; ``naturally I had something on my mind, then boom, I am reading what it is writing,'' P00348, letter/co-writing) and that completions visually covered their own text (15\% vs.\ 3\%). 

Scaffolding's problems were instead about presence: 16\% of scaffolding participants (vs.\ 11\% in co-writing) reported that the assistant was unavailable, showed error messages, or never appeared (``It said it was `offline' for my whole writing session,'' P00140, reflection/scaffolding), and 15\% (vs.\ 8\%) said they did not need or barely used it: ``I got so into writing it from my own past experiences and heart, that the words flowed without anything else needed'' (P00057, letter/scaffolding). Disruption was not unique to co-writing, however; P00470 (reflection/scaffolding) described how reading each new question triggered another pause and another question until ``it would snowball,'' adding, ``Oftentimes I would rather think on me own if I am stuck.''

\subsubsection{Authenticity and Voice in Emotional Writing}
A smaller group of participants (8\%; 10\% in co-writing, 5\% in scaffolding) raised concerns about authenticity and voice that were specific to the emotional nature of the task. P00231 (letter/co-writing) found the text suggestions ``more fake and inauthentic. I feel that the writing needed to be genuine and sincerely written in my own words,'' and P00459 (reflection/co-writing) doubted that prediction could work ``for these kind've emotional topics \ldots especially since it's coming from the heart.'' P00094 (letter/scaffolding) declined to use the tool because ``AI tends to erase personality and authenticity from writing so if it's something potentially meaningful, I just would prefer not to use AI tools,'' and P00019 (letter/scaffolding) noticed a drift in the other direction: ``I also got less involved as the process went on and it just felt like I was completely relying on the ai.'' 

\subsubsection{Design Requests}
Consistent with these concerns, participants' design requests centered on control and quality rather than on more generated text: better and less repetitive suggestions (15\%; ``If something is rejected, it shouldn't reappear,'' P00053, reflection/co-writing), control over when the AI appears (7\%; ``I wish I could not see it until I knew I needed it and could then click a button to open it,'' P00315, letter/scaffolding), a choice among alternatives (``Instead of providing one example sentence, provide three to choose between,'' P00250, letter/co-writing), adaptation to the writer's own style (``giving users more control over the tone, level of detail, and writing style,'' P00220, letter/co-writing), and, for a few, a conversational mode (``a chatbot option would be nice. So users can ask questions and recommendations directly,'' P00448, reflection/scaffolding). 

\section{Discussion}
Our study examined how gratitude writing format and AI-assistance mode shape well-being outcomes (RQ1), the writing experience and perceived agency (RQ2), and self-disclosure (RQ3). Prior work has shown that gratitude writing interventions can improve well-being, but that their effects vary depending on intervention format, effort, and person-activity fit \cite{davis2016thankful,dickens2017using,diniz2023effects,lyubomirsky2011becoming,lyubomirsky2013simple}. At the same time, research on AI-assisted writing has demonstrated that generative AI assistance can facilitate ideation and text production while also shaping writers' expression, cognitive investment, agency, and ownership \cite{dhillon2024shaping,draxler2024ai,gero2022sparks,lee2022coauthor}. Our findings bring these two bodies of work together and show that the role that AI takes in gratitude writing matters differently depending on whether we consider psychological outcomes, the experience of writing, or the content writers disclose.

\subsection{RQ1: Gratitude Writing Format and AI-assistance Mode Shape Different Aspects of Well-being}
For RQ1, most well-being measures improved from pre- to post-writing across conditions, but the pattern of change differed by writing and AI-assistance type. Gratitude reflection produced larger gains in life satisfaction, whereas co-writing produced larger gains in positive affect and affect balance. These findings are consistent with prior gratitude research that showed both reflective and interpersonal gratitude activities can promote positive outcomes, while their effects vary across intervention formats and outcomes \cite{davis2016thankful,dickens2017using,regan2023some,toepfer2012letters}. Our results extend this work by suggesting that the way AI participates and collaborates in the writing process may also shape some immediate affective outcomes.

At the same time, greater engagement with AI was more consistently associated with interface experience than with well-being change. Thus, interacting more with AI, or accepting more of its suggestions, should not necessarily be interpreted as greater psychological benefit. This distinction is particularly important for well-being writing since generating, selecting, and articulating meaningful experiences may themselves be part of the intervention process. However, we cannot conclude whether AI assistance itself produced the overall pre-to-post improvements in well-being since our study did not include a condition without AI assistance. Rather, our findings suggest that both AI-assistance modes were compatible with the short-term outcomes of gratitude writing while shaping particular outcomes differently.

\subsection{RQ2: AI Writing Assistance May Change Reflective Processes}
For RQ2, our findings reveal a trade-off between the usability of AI assistance and writers' sense of agency. Co-writing received higher usability and user-experience ratings, whereas scaffolding resulted in stronger human-ownership attribution. Moreover, within co-writing, greater suggestion uptake was associated with lower perceived ownership, control, and human contribution. This finding extends prior research that showed direct AI contributions can create discrepancies between authorship and ownership \cite{draxler2024ai}, steer writers toward AI-salient language \cite{arnold2020predictive,jakesch2023co}, and shift writers from generating ideas toward evaluating AI-generated alternatives \cite{bhat2026reactive}. Prior work has similarly found trade-offs among the level of AI assistance, cognitive investment, and perceived ownership \cite{dhillon2024shaping}. Our study results extend this work by showing why this trade-off may be particularly consequential in gratitude writing.

Unlike many productivity-oriented writing tasks, gratitude writing is not solely concerned with the quality or efficiency of the resulting text. The act of identifying a meaningful experience, considering why it matters, and finding one's own words to express it is itself part of the intended activity. Prior work on expressive and gratitude writing similarly conceptualizes writing as a process through which people organize experiences, elaborate on their meaning, and direct attention toward personally significant events and relationships \cite{pennebaker1997writing,frattaroli2006experimental,emmons2003counting}. Direct text generation may therefore alter not only authorship of the final artifact, but also who performs the reflective work leading to that artifact.

Participants' qualitative feedback illustrates this distinction. Co-writing could make expressions easier and help participants continue when they were unsure what to write, but some participants also described suggestions as interrupting their own thought process or making emotionally meaningful writing feel less genuine. Scaffolding, by contrast, left participants responsible for translating reflection into language, although its prompts could be overlooked or interrupt moments of quiet thought.

Thus, for well-being writing, the relevant design question may be less about maximizing perceived usability or minimizing effort and more about determining which parts of the activity should remain effortful and self-generated because they contribute to reflection itself. 

\subsection{RQ3: Writing Format Matters More Than AI Support for Self-Disclosure}
For RQ3, writing format shaped the nature of self-disclosure, whereas AI-assistance mode did not significantly affect disclosure depth. Gratitude letters contained deeper feelings-based disclosure, while gratitude reflections contained deeper disclosure of personal thoughts. This pattern aligns with the different structure of these established gratitude interventions. Gratitude reflections center private consideration of positive experiences and circumstances, while gratitude letters require writers to formulate appreciation in relation to a particular person and potential audience \cite{emmons2003counting,seligman2005positive,regan2023some,toepfer2012letters}.

This finding is notable given prior work showing that AI suggestions can influence what writers consider and ultimately produce \cite{arnold2020predictive,buschek2021impact,jakesch2023co}. In our study, however, whether AI provided reflective directions or directly generated language did not significantly change the measured depth of participants' disclosure. Instead, the gratitude writing format more strongly shaped the type of personal content participants disclosed: letters elicited deeper feelings-based disclosure, while reflections elicited deeper thought-based disclosure. This suggests that the type of gratitude writing task may play a larger role than the type of AI assistance in shaping what people disclose.

\subsection{Designing AI to Support Without Taking Over}
Across the three RQs, our findings suggest that effective AI support for well-being writing should not be defined simply by maximizing AI engagement or reducing writing effort. Rather than treating scaffolding and co-writing as fixed alternatives, future systems could allow users to regulate the timing and level of AI involvement. For example, a system could allow users to choose between receiving reflective questions, getting help generating ideas, or receiving suggested wording, depending on the type of support they need. Such designs could preserve the benefits identified in prior work (e.g., overcoming writing blocks and supporting idea development \cite{gero2022sparks,lee2022coauthor,wan2024felt}), while reducing the risk that AI-generated content replaces participants' own process of reflecting on and expressing meaningful experiences. More broadly, our results highlight the importance of matching the form and timing of AI assistance to the goals of the writing activity and the writer's needs. 

In addition, designers should be cautious about optimizing for suggestion acceptance or the amount of AI-generated text incorporated into a response. In our study, greater uptake of co-writing suggestions was associated with weaker perceived agency, while AI engagement was more closely related to interface experience than to well-being outcomes. For well-being writing, the goal may therefore be to provide the minimum assistance needed to sustain users' own reflection, rather than maximizing AI contribution.

Several limitations motivate future work. Our study did not include a no-AI condition, and thus we cannot isolate the contribution of AI to the observed well-being changes. The scaffolding and co-writing conditions also differed in timing, presentation, and type of assistance, which makes it difficult to attribute differences to a single design feature. Finally, our study examined a single short writing session and used an LLM-based proxy measure of self-disclosure. Future studies should explore repeated use, validate disclosure measures with human coding, and investigate adaptive forms of AI support that balance assistance with users' agency, authenticity, and ownership over time.

\section{Conclusion}
In this study, we investigated how different forms of AI assistance and gratitude writing activities shape people's well-being, writing experience, and self-disclosure. Our findings suggest that neither scaffolding nor co-writing is universally preferable. Co-writing provided more immediate and usable support and was associated with larger gains in positive affect and affect balance, but greater reliance on its suggestions was also associated with weaker perceived agency. Scaffolding better preserved human ownership by leaving participants responsible for articulating their experiences, although its support was sometimes less noticeable. Meanwhile, the gratitude writing format, rather than the AI-assistance mode, more strongly shaped the nature of writers' self-disclosure, with letters eliciting deeper feelings and reflections eliciting deeper thoughts. Together, these findings highlight an important consideration for AI-supported well-being writing: AI assistance should support the process of reflection and expression without unnecessarily performing that work for the user. Rather than maximizing AI contribution, future systems could provide context-aware support that infers users' needs during writing and dynamically adjusts when and how support is provided. Such systems could distinguish between moments when users are productively reflecting and moments when they need help generating ideas or expressing their thoughts, and provide only the level of assistance needed to sustain the writing process. In doing so, AI writing tools may help people engage in personally meaningful writing while preserving their active role in the reflective process.

\section*{Ethics and Privacy Statement}
AI-supported gratitude writing may make reflective guidance more accessible, but it also introduces privacy and autonomy risks because users may disclose sensitive personal and relational information, and our findings associate greater engagement with co-writing suggestions with weaker perceived agency. Systems should therefore clearly disclose how users' writing is processed, stored, and reused; minimize the collection and secondary use of emotionally sensitive data; and preserve users' control over whether AI-generated content enters or remains in their writing. These findings should not be used to treat suggestion uptake as evidence of psychological benefit, infer users' mental-health status for consequential decisions, or position AI-supported writing as a substitute for professional care.

\bibliographystyle{ACM-Reference-Format}
\bibliography{ref}

\newpage

\appendix
\section{Measures}

\label{app:measures}

We administer measures of life satisfaction, gratitude, affect, and moral elevation immediately before and after the writing activity. Participants also post-task measures of writing self-efficacy, user experience, usability, ownership, control, leadership, and perceived text match. We additionally derive behavioral engagement measures from system logs and use LLM-generated ratings to characterize self-disclosure in participants' final texts.

\subsection{Well-Being Outcomes}

\paragraph{Satisfaction With Life.}

We measure life satisfaction using the five-item Satisfaction With Life Scale (SWLS) \cite{diener1985satisfaction}. Participants rate each statement on a seven-point scale from 1 (\emph{strongly disagree}) to 7 (\emph{strongly agree}). We sum the five items to obtain scores ranging from 5 to 35, with higher scores indicating greater life satisfaction. A score is computed only when all five items are available.

\paragraph{Trait Gratitude.}

We assess dispositional gratitude using the six-item Gratitude Questionnaire (GQ-6) \cite{mccullough2002grateful}. Participants respond on a seven-point scale from 1 (\emph{strongly disagree}) to 7 (\emph{strongly agree}). We reverse-score Items 3 and 6 using $8-x$ and sum all items, producing scores from 6 to 42; higher scores indicate a stronger tendency to experience gratitude. A score is computed only when all six items are available.

\paragraph{Positive and Negative Affect.}

We assess affect using the 12-item Scale of Positive and Negative Experience (SPANE) \cite{diener2010new}. Participants rate six positive and six negative feelings on a five-point scale. Responses use intensity anchors from 1 (\emph{not at all}) to 5 (\emph{very much}). We map anchor formats onto the same 1--5 metric and treat them as comparable in the primary analyses. We sum Items 1, 3, 5, 7, 10, and 12 to calculate SPANE-P and Items 2, 4, 6, 8, 9, and 11 to calculate SPANE-N, each ranging from 6 to 30. We then subtract SPANE-N from SPANE-P to calculate affect balance (SPANE-B), ranging from $-24$ to 24; higher scores indicate a more positive affective balance. Each component is computed only when all six constituent items are available.

\paragraph{Moral Elevation.}

We assess state moral elevation using six items adapted from Schnall et al.~\cite{schnall2010elevation}. The items capture feeling moved, uplifted, optimistic about humanity, warmth in the chest, motivated to help others, and motivated to become a better person. Participants rate each experience from 1 (\emph{not at all}) to 9 (\emph{very strongly}). We average Items 1--6, with higher scores indicating stronger moral elevation; the separately administered seventh item (\emph{happy}) is not included. A score is computed only when all six scored items are available.

\subsection{Writing Experience}

\paragraph{Writing Self-Efficacy.}

We assess participants' confidence in writing using nine items adapted from the Nutrition and Physical Exercise Self-Efficacy Scales \cite{schwarzer2009health}. We replace references to nutrition and exercise with writing-specific language. Participants respond on a four-point certainty scale from 1 (\emph{very uncertain}) to 4 (\emph{very certain}). Four writing-process items assess participants' confidence in developing ideas, revising repeatedly, rethinking their approach, and writing without extensive support. Five writing-barrier items assess confidence in carrying out writing intentions while worried, depressed, tense, tired, or busy. We compute complete-item sums for writing-process efficacy (range 4--16), writing-barrier efficacy (range 5--20), and overall writing self-efficacy across all nine items (range 9--36); higher scores indicate greater writing self-efficacy. The adapted version is presented in \Cref{tab:writing-self-efficacy-adaptation}.

\paragraph{User Experience and Usability.}

We measure overall user experience using a 21-item version of the User Experience Questionnaire (UEQ) \cite{laugwitz2008construction}. The UEQ uses seven-point semantic-differential items covering attractiveness, perspicuity, efficiency, dependability, stimulation, and novelty. Because some items place the positive pole on the left, we reverse-score Items 3, 4, 5, 8, 9, 11, 16, 17, 18, and 20 using $8-x$ so that higher values consistently indicate a more positive evaluation. We then average the complete set of 21 items on the original 1--7 scale; a score is computed only when all 21 items are available. We assess perceived usability using the 10-item System Usability Scale (SUS) \cite{lewis2018system}. Participants respond on five-point agreement scales. Following the standard SUS procedure, we subtract 1 from each odd-numbered item, subtract each even-numbered item from 5, sum the ten contributions, and multiply by 2.5 to obtain a total score from 0 to 100; higher scores indicate greater perceived usability. A SUS score is computed only when all ten items are available.

\paragraph{Ownership, Control, and Text Match.}

We adapt measures of perceived authorship, ownership, and control from \cite{draxler2024ai}, replacing references to postcards with the more general term \emph{text} and removing items that do not apply to the present writing task. Participants rate all non-slider items on seven-point agreement scales.

The five-item ownership measure assesses whether participants view themselves as the primary contributor, drafter, approver, and author of the text. No ownership items are reverse-scored. The seven-item control measure assesses intentionality, responsibility for the resulting text, control over the writing process, perceptions of the AI as a ghostwriter or controllable tool, and freedom to determine the writing approach and organization, with reversing the items as in \cite{draxler2024ai}. We then average each complete scale so that higher scores indicate greater perceived ownership or control.

The four-item text-match measure assesses whether participants would send the text to friends or family, are satisfied and happy with the resulting text, and are pleased with its flow. We average the items, with higher scores indicating a stronger perceived match between the participant and the resulting text. Ownership, control, and text-match scores are each computed only when all of their constituent items are available.

Participants separately indicate ownership, control, and leadership using three continuous Me--AI sliders adapted from \cite{draxler2024ai}: ``To whom should this text belong?'', ``Who was in control of the content?'', and ``Who took the lead in writing during this interaction method?'' Raw slider values range from $+100$ (\emph{Me}) to $-100$ (\emph{AI}). For inferential analyses, we multiply each raw score by $+1$ so that higher values indicate greater attribution to the participant and lower values indicate greater attribution to the AI. The three sliders are analyzed as standalone outcomes and are not averaged into the Likert ownership or control composites.

Full adapted measures we used are shown in \Cref{tab:control-agency-adaptation,tab:leadership-adaptation,tab:ownership-adaptation,tab:text-match-adaptation}.

\subsection{Behavioral Engagement}

We derive engagement measures from the writing interface logs. For co-writing, we record the number of autocomplete suggestions shown, accepted, and rejected. The primary co-writing engagement measure is the resolved acceptance rate,

$$\frac{\text{accepted}}{\text{accepted}+\text{rejected}},
$$

computed among participants who received at least one autocomplete suggestion. For scaffolding, we record like and dislike responses to AI prompts. The primary scaffolding engagement measure is the like rate,
$$\frac{\text{likes}}{\text{likes}+\text{dislikes}},
$$
computed among participants who rated at least one scaffolding prompt. We also record the final submitted word count for each participant.

\subsection{LLM-Rated Self-Disclosure}

To characterize the content of participants' writing, we score each final text on three self-disclosure dimensions using GPT-5.2-chat with prompts adapted from prior self-disclosure coding schemes: feelings, information, and thoughts. Each dimension is rated on a three-point depth scale from 1 to 3, where higher scores indicate deeper disclosure. Feelings capture the emotional depth of the writing, information captures the depth of personal factual disclosure, and thoughts capture the depth of personal ideas or reflections. When all three dimension scores are available, we compute an overall disclosure mean as their average. These LLM ratings are treated as proxy annotations of disclosure content rather than as human-coded writing-quality scores.

\subsection{Open-Ended Feedback}

Following the quantitative measures, participants respond to four open-ended questions about their experience:

\begin{enumerate}

\item ``What did you like about the system?''

\item ``What did you not like about the system?''

\item ``Please share anything that can be improved about the system.''

\item ``Please share any other thoughts or feedback you have about the writing experience.''

\end{enumerate}

We use these responses to contextualize participants' quantitative ratings and identify perceived benefits, limitations, and opportunities for improving the AI-supported writing experience. For exploratory qualitative analyses, we apply dictionary-assisted, multi-label thematic coding to the open-ended responses and report theme frequencies among respondents to each question.

\begingroup
\scriptsize
\setlength{\tabcolsep}{4pt}
\renewcommand{\arraystretch}{1.12}

\begin{table*}[tp]
\caption{Adaptation of the health-specific self-efficacy items to measure writing self-efficacy.}
\label{tab:writing-self-efficacy-adaptation}
\centering
\begin{tabular}{@{}>{\RaggedRight\arraybackslash}p{0.32\textwidth} >{\RaggedRight\arraybackslash}p{0.35\textwidth} >{\RaggedRight\arraybackslash}p{0.27\textwidth}@{}}
\toprule
\textbf{Health-specific self-efficacy source} &
\textbf{Adapted writing self-efficacy item} &
\textbf{Modification} \\
\midrule
How certain are you that you could overcome the following barriers?

\textbf{The Nutrition Self-Efficacy Scale}

I can manage to stick to healthful foods \ldots
&
The following questions are about your confidence in writing.

How certain are you that you could overcome the following barriers to writing?

I can manage to write \ldots
&
Added writing context to the prompt; changed ``stick to healthful foods'' to ``write.'' \\
\midrule
1. Even if I need a long time to develop the necessary routines.
& 1. Even if I need a long time to develop my ideas.
& Changed ``the necessary routines'' to ``my ideas.'' \\
2. Even if I have to try several times until it works.
& 2. Even if I have to write several times until it works.
& Changed ``try'' to ``write.'' \\
3. Even if I have to rethink my entire way of nutrition.
& 3. Even if I have to rethink my entire way of writing.
& Changed ``nutrition'' to ``writing.'' \\
4. Even if I do not receive a great deal of support from others when making my first attempts.
& 4. Even if I do not receive a great deal of support from others when making my first attempts.
& No change. \\
5. Even if I have to make a detailed plan.
& \textemdash
& Removed. \\
\midrule
\textbf{The Physical Exercise Self-Efficacy Scale}

How certain are you that you could overcome the following barriers?

I can manage to carry out my exercise intentions \ldots
&
I can manage to carry out my writing intentions \ldots
&
Did not repeat the first half of the prompt; changed ``exercise'' to ``writing.'' \\
\midrule
1. Even when I have worries and problems.
& 1. Even when I have worries and problems.
& No change. \\
2. Even if I feel depressed.
& 2. Even if I feel depressed.
& No change. \\
3. Even when I feel tense.
& 3. Even when I feel tense.
& No change. \\
4. Even when I am tired.
& 4. Even when I am tired.
& No change. \\
5. Even when I am busy.
& 5. Even when I am busy.
& No change. \\
\bottomrule
\end{tabular}
\end{table*}

\begin{table*}[tp]
\caption{Adaptation of the text-match items.}
\label{tab:text-match-adaptation}
\centering
\begin{tabular}{@{}>{\RaggedRight\arraybackslash}p{0.32\textwidth} >{\RaggedRight\arraybackslash}p{0.35\textwidth} >{\RaggedRight\arraybackslash}p{0.27\textwidth}@{}}
\toprule
\textbf{Source text-match item} &
\textbf{Adapted text-match item} &
\textbf{Modification} \\
\midrule
I would actually send this postcard to my friends and family.
& I would actually send this text to my friends and family.
& Changed ``postcard'' to ``text.'' \\
I would have written a similar postcard by myself.
& \textemdash
& Removed. \\
The postcard mostly contains words and/or phrases that I usually use when writing in English.
& \textemdash
& Removed. \\
I am satisfied with the text.
& I am satisfied with the text.
& No change. \\
\textemdash
& I am happy with the text produced by this process.
& Added. \\
\textemdash
& I am pleased with the flow of the text.
& Added. \\
\bottomrule
\end{tabular}
\end{table*}

\endgroup

\subsection{Number of Participants for Each Measure}
\label{app:numpartcipant}

\begin{table*}[t]
\centering
\caption{Number of participants included in each measure (complete-case).}
\label{tab:measure-ns}
\scriptsize
\setlength{\tabcolsep}{4pt}
\renewcommand{\arraystretch}{0.95}
\begin{tabular}{llrr}
\toprule
Domain & Measure & $N$ & \% of analytic sample \\
\midrule
Sample & Analytic sample & 401 & 100.0 \\
\midrule
\multirow{6}{*}{Baseline (pre)}
 & Life satisfaction (SWLS) & 399 & 99.5 \\
 & Gratitude (GQ-6) & 399 & 99.5 \\
 & Positive affect (SPANE-P) & 396 & 98.8 \\
 & Negative affect (SPANE-N) & 399 & 99.5 \\
 & Affect balance (SPANE-B) & 394 & 98.3 \\
 & Moral elevation & 399 & 99.5 \\
\midrule
\multirow{6}{*}{Well-being (pre--post)}
 & Life satisfaction (SWLS) & 384 & 95.8 \\
 & Gratitude (GQ-6) & 379 & 94.5 \\
 & Positive affect (SPANE-P) & 378 & 94.3 \\
 & Negative affect (SPANE-N) & 385 & 96.0 \\
 & Affect balance (SPANE-B) & 376 & 93.8 \\
 & Moral elevation & 376 & 93.8 \\
\midrule
\multirow{11}{*}{Writing experience}
 & Writing efficacy: total & 399 & 99.5 \\
 & Writing efficacy: process & 399 & 99.5 \\
 & Writing efficacy: barrier & 400 & 99.8 \\
 & Likert ownership & 380 & 94.8 \\
 & Likert control & 380 & 94.8 \\
 & Text match & 383 & 95.5 \\
 & UEQ & 375 & 93.5 \\
 & SUS & 379 & 94.5 \\
 & Ownership slider & 383 & 95.5 \\
 & Leadership slider & 383 & 95.5 \\
 & Control slider & 383 & 95.5 \\
\midrule
\multirow{2}{*}{Engagement}
 & Co-writing acceptance rate (shown $>0$) & 199 & 49.6 \\
 & Scaffolding like rate (rated $\ge 1$) & 110 & 27.4 \\
\midrule
\multirow{2}{*}{Text}
 & Final word count & 401 & 100.0 \\
 & LLM self-disclosure (feelings, info, thoughts) & 401 & 100.0 \\
 \midrule
 \multirow{4}{*}{Open-ended feedback}
 & Like: What did you like about the system? & 374 & 93.3 \\
 & Dislike: What did you not like about the system?  & 355 & 88.5 \\
 & Improvements: What can be improved about the system?  & 297 & 74.1 \\
 & Others: Other thoughts about the writing experience  & 215 & 53.6 \\
\bottomrule
\end{tabular}
\end{table*}

\begin{table*}[tp]
\caption{Adaptation of the sense-of-ownership items. The source items assess perceived authorship and ownership using criteria from the International Committee of Medical Journal Editors and Nylenna et al. (2014).}
\label{tab:ownership-adaptation}
\centering
\begin{tabular}{@{}>{\RaggedRight\arraybackslash}p{0.32\textwidth} >{\RaggedRight\arraybackslash}p{0.35\textwidth} >{\RaggedRight\arraybackslash}p{0.27\textwidth}@{}}
\toprule
\textbf{Source ownership item} &
\textbf{Adapted ownership item} &
\textbf{Modification} \\
\midrule
I am the main contributor to the content of this postcard.
& I am the main contributor to the content of this text.
& Changed ``postcard'' to ``text.'' \\
I have made substantial contributions to the content of this postcard.
& I have made substantial contributions to the content of this text.
& Changed ``postcard'' to ``text.'' \\
I have drafted this postcard.
& I drafted this text.
& Changed ``postcard'' to ``text'' and removed ``have.'' \\
I revised this postcard critically.
& \textemdash
& Removed. \\
I have given final approval of this text being uploaded.
& I have given final approval for this text being uploaded.
& Changed ``of'' to ``for.'' \\
I am accountable for all aspects of the text.
& \textemdash
& Removed. \\
I am responsible for at least part of the text.
& \textemdash
& Removed. \\
My name should appear underneath this postcard.
& \textemdash
& Removed. \\
I feel like I am the author of this text.
& I feel like I am the author of the text.
& Changed ``postcard'' to ``text'' and ``this'' to ``the.'' \\
To whom should this text belong? (Slider: Me--AI)
& To whom should this text belong? (Slider: Me--AI)
& No change. \\
\bottomrule
\end{tabular}
\end{table*}

\begin{table*}[tp]
\caption{Adaptation of the sense-of-leadership items.}
\label{tab:leadership-adaptation}
\centering
\begin{tabular}{@{}>{\RaggedRight\arraybackslash}p{0.32\textwidth} >{\RaggedRight\arraybackslash}p{0.35\textwidth} >{\RaggedRight\arraybackslash}p{0.27\textwidth}@{}}
\toprule
\textbf{Source leadership item} &
\textbf{Adapted leadership item} &
\textbf{Modification} \\
\midrule
I felt like I was writing the text and the artificial intelligence was assisting me.
& \textemdash
& Removed. \\
I felt like the artificial intelligence was writing the text and I was assisting.
& \textemdash
& Removed. \\
Who took the lead in writing during this interaction method? (Slider: Me--AI)
& Who took the lead in writing during this interaction method? (Slider: Me--AI)
& Grouped with the other slider questions. \\
\bottomrule
\end{tabular}
\end{table*}

\begin{table*}[tp]
\caption{Adaptation of the sense-of-control and sense-of-agency items.}
\label{tab:control-agency-adaptation}
\centering
\begin{tabular}{@{}>{\RaggedRight\arraybackslash}p{0.32\textwidth} >{\RaggedRight\arraybackslash}p{0.35\textwidth} >{\RaggedRight\arraybackslash}p{0.27\textwidth}@{}}
\toprule
\textbf{Source control or agency item} &
\textbf{Adapted control item} &
\textbf{Modification} \\
\midrule
Who was in control of the content? (Slider: Me--AI)
& Who was in control of the content? (Slider: Me--AI)
& Grouped with the other slider questions. \\
\midrule
\multicolumn{3}{@{}l}{\textbf{Sense of Agency Subscale}} \\
\midrule
I was just an instrument in the hands of the AI system.
& \textemdash
& Removed. \\
The resulting text just happened without my intention.
& The resulting text just happened without my intention.
& No change. \\
I was the author of my actions while interacting with the system.
& \textemdash
& Removed. \\
The consequences of my actions felt like they don't logically follow my actions while interacting with the system.
& \textemdash
& Removed. \\
The outcomes of my actions generally surprised me.
& \textemdash
& Removed. \\
While I am interacting with the system, I feel like I am a remote controlled robot.
& \textemdash
& Removed. \\
I am completely responsible for the resulting text.
& I am completely responsible for the resulting text.
& No change. \\
\midrule
\multicolumn{3}{@{}l}{\textbf{Sense of Control}} \\
\midrule
It felt like I was in control of the text during the task.
& I felt in full control of my writing.
& Rewrote item. \\
I felt like the AI system was acting as a ghostwriter, writing the postcard on my behalf.
& I felt like the AI system was acting as a ghostwriter, writing the text on my behalf.
& Changed ``postcard'' to ``text.'' \\
I felt like the AI system was acting as a tool which I could control.
& I felt like the AI system was acting as a tool which I could control.
& No change. \\
\textemdash
& I was able to decide how to approach my writing tasks.
& Added. \\
\textemdash
& I felt free to organize my ideas in the way that made the most sense.
& Added. \\
\bottomrule
\end{tabular}
\end{table*}

\begin{table*}[tp]
\caption{Adaptation of the text-match items.}
\label{tab:text-match-adaptation}
\centering
\begin{tabular}{@{}>{\RaggedRight\arraybackslash}p{0.32\textwidth} >{\RaggedRight\arraybackslash}p{0.35\textwidth} >{\RaggedRight\arraybackslash}p{0.27\textwidth}@{}}
\toprule
\textbf{Source text-match item} &
\textbf{Adapted text-match item} &
\textbf{Modification} \\
\midrule
I would actually send this postcard to my friends and family.
& I would actually send this text to my friends and family.
& Changed ``postcard'' to ``text.'' \\
I would have written a similar postcard by myself.
& \textemdash
& Removed. \\
The postcard mostly contains words and/or phrases that I usually use when writing in English.
& \textemdash
& Removed. \\
I am satisfied with the text.
& I am satisfied with the text.
& No change. \\
\textemdash
& I am happy with the text produced by this process.
& Added. \\
\textemdash
& I am pleased with the flow of the text.
& Added. \\
\bottomrule
\end{tabular}
\end{table*}

\subsection{Gratitude Writing Tasks}
\label{app:writing_prompt}

\noindent\textbf{Gratitude Letter Prompt.}
\begin{lstlisting}[style=llmprompt]
Please take a moment to think back over the past several years of your life and remember an instance when someone did something for which you are extremely grateful. Think of the people -- parents, children, spouses/partners, relatives, friends, neighbors, teachers, employers, and so on -- who have been especially generous and thoughtful towards you. For example, you might think of a parent who sacrificed to support your education, a friend who was there for you during a difficult time, a teacher who inspired you to pursue your passion, or a mentor who helped you navigate a career decision. Now, for the next 8 minutes, write a letter to one of these individuals. Use the instructions below to help guide you through this process:
* Think of a specific person: a parent, grandparent, friend, teacher, coach, neighbor, or mentor who made a meaningful difference in your life.
* Write as though you are directly addressing them (e.g., "Dear Mom," or "Dear Professor Johnson,").
* Describe in specific terms the kind act they did for you and how it affected your life.
* Describe what you are doing now and how you often remember their efforts.
* Do not worry about perfect grammar and spelling. Anything you write will remain strictly confidential.
\end{lstlisting}

\noindent\textbf{Personal Gratitude Reflection Prompt.}
\begin{lstlisting}[style=llmprompt]
Write about the many things in your life, both large and small, that you have to be grateful about. These might include particular supportive relationships, sacrifices or contributions that others have made for you, facts about your life such as your advantages and opportunities, or even gratitude for life itself, and the world that we live in. Outline these reasons in as much detail as you can.
* What are the small things in daily life you are grateful for? (e.g., a warm cup of coffee in the morning, a favorite song, a sunny day)
* Who are the people who have supported you? (e.g., a parent who always encouraged you, a friend who listened when you needed to talk, a teacher who believed in you)
* What opportunities or advantages have you had? (e.g., access to education, a safe place to live, a meaningful job or hobby)
\end{lstlisting}

\section{LLM Prompts for AI Support}
\label{app:llm_prompts}
The system used three LLM request types: reflective questions in scaffolding, on-demand suggestions in co-writing, and inline completions in co-writing. Requests were sent to an Azure OpenAI deployment (\emph{gpt-4o}) through the Chat Completions API using API version \texttt{2024-02-15-preview}. Scaffolding questions and on-demand co-writing suggestions used a temperature of $0.8$ and a maximum of 280 output tokens. Inline completions used a temperature of $0.4$ and a maximum of 60 output tokens.

Before being inserted into a prompt, each draft was sanitized by removing empty lines; lines beginning with \texttt{system:}, \texttt{assistant:}, or \texttt{user:}; and lines containing selected prompt-injection terms, including \texttt{ignore all previous}, \texttt{override}, \texttt{forget}, and \texttt{bypass}. The resulting draft was truncated to 2,000 characters. Empty drafts were represented as \texttt{No text yet.}. In the templates below, angle-bracketed terms denote values inserted at runtime rather than literal prompt text.
\subsection{Scaffolding Prompt}
\label{app:llm_scaffolding}

The scaffolding condition generated a single reflective question based on the participant’s current draft. Previously generated questions were included when available to discourage repetition.

\noindent\textbf{System Prompt.}
\begin{lstlisting}[style=llmprompt]
You are a writing coach for a gratitude study. Your role is to help participants reflect on gratitude by asking insightful questions. Each question you ask must be substantially different from any you have already asked.
\end{lstlisting}

\noindent\textbf{User Prompt.}
\begin{lstlisting}[style=llmprompt]
A participant is writing a <GRATITUDE\_TYPE>.
Their current draft is below:

<CURRENT\_DRAFT>

<OPTIONAL\_HISTORY\_BLOCK>
Based on their draft, provide one short, reflective question to encourage them to write more. Ask about something NEW and DIFFERENT from any previous questions. If the draft is not about gratitude, your question should gently guide them back to that topic.
\end{lstlisting}

When previous questions were available, \texttt{<OPTIONAL\_HISTORY\_BLOCK>} was replaced with the following block; otherwise, it was omitted.

\begin{lstlisting}[style=llmprompt]
You have already asked these questions (DO NOT repeat or rephrase them):
<NUMBERED\_PREVIOUS\_QUESTIONS>
\end{lstlisting}

Here, \texttt{<GRATITUDE\_TYPE>} was either \texttt{reflection} or \texttt{gratitude\_letter}, \texttt{<CURRENT\_DRAFT>} contained the sanitized draft, and \texttt{<NUMBERED\_PREVIOUS\_QUESTIONS>} contained the previously generated questions as a numbered list.
\subsection{Co-Writing Prompts}
\label{app:llm_cowriting}

The co-writing condition offered two forms of assistance: on-demand writing suggestions through the \emph{Ask Ami} button and automatic inline completions.

\subsubsection{On-Demand Suggestions}

When participants selected \emph{Ask Ami}, the system generated one or two concrete suggestions for making the current draft more heartfelt and detailed.

\noindent\textbf{System Prompt.}
\begin{lstlisting}[style=llmprompt]
You are a writing coach for a gratitude study. Your role is to help participants improve their gratitude writing by providing concrete suggestions.
\end{lstlisting}

\noindent\textbf{User Prompt.}
\begin{lstlisting}[style=llmprompt]
A participant is writing a <GRATITUDE_TYPE>.
Their current draft is below:

<CURRENT_DRAFT>

Based on their draft, provide 1-2 specific, numbered suggestions to make the writing more heartfelt and detailed. If the draft is not about gratitude, your suggestions should gently guide them back to that topic.
\end{lstlisting}

Here, \texttt{<GRATITUDE\_TYPE>} and \texttt{<CURRENT\_DRAFT>} were populated as described above.

\subsubsection{Inline Completions}

Inline completions were generated from the sanitized text preceding the participant’s cursor. The system did not request a completion when this text contained fewer than 10 characters or was represented as \texttt{No text yet.}. The writing type was expressed as either \emph{personal gratitude reflection} or \emph{gratitude letter} in the system prompt.

\noindent\textbf{System Prompt.}
\begin{lstlisting}[style=llmprompt]
You are a quiet, unobtrusive writing assistant. Someone is writing a <WRITING_TYPE>.

Rules:

* Complete their current sentence or thought naturally. Only provide the continuation text - never repeat what they wrote.
* Keep completions SHORT (5-15 words). Less is more.
* Mirror their EXACT voice: use the same vocabulary level, sentence length, and emotional register they are using. If they write simply, you write simply. If they are calm, be calm. Never elevate or embellish their tone.
* Do NOT add dramatic language, superlatives, or flowery phrasing (e.g. avoid "truly", "deeply", "profoundly", "incredible", "amazing").
* Sound like the next few words THEY would naturally type, not like an AI.
* Stay grounded in the specific details they mention - do not generalize or abstract.
\end{lstlisting}

\noindent\textbf{User Prompt.}
\begin{lstlisting}[style=llmprompt]
Continue this text naturally:

<TEXT_BEFORE_CURSOR>
\end{lstlisting}

Here, \texttt{<WRITING\_TYPE>} was either \texttt{personal gratitude reflection} or \texttt{gratitude letter}, and \texttt{<TEXT\_BEFORE\_CURSOR>} contained the sanitized text preceding the cursor. The returned completion was trimmed, and the system inserted a leading space when necessary to separate it from the participant’s existing text.

\section{LLM Self-Disclosure Scoring Prompts}
\label{app:llm_disclosure_prompt}

Each final writing text is scored independently on three self-disclosure dimensions with GPT-5.2-chat 5 times.
The dimension-specific instructions below are sent as the \emph{system} message.
The \emph{user} message is identical across dimensions:

\begin{lstlisting}[style=llmprompt]
Score this text. Reply with ONLY the integer 1, 2, or 3.

TEXT:
<participant final writing>
\end{lstlisting}

The model is instructed to return only the integer score $1$, $2$, or $3$.
Texts longer than $12{,}000$ characters are truncated to the first and last $6{,}000$ characters with a truncation marker.

\noindent\textbf{Feelings.}
\begin{lstlisting}[style=llmprompt]
As a social scientist, your task is to annotate the depth of this self-disclosure statements.
Evaluate the feelings of the self-disclosure statement:
    - Score 1: No expressing of feelings at all. The speech may include a prosaic description of facts or personal ideas, without expressing any emotions or affective relevance
    - Score 2: Expressing of some mild feelings
    - Score 3: Expressions of deep feelings
Only give the score (number) without any explanation
\end{lstlisting}

\noindent\textbf{Information.}
\begin{lstlisting}[style=llmprompt]
As a social scientist, your task is to annotate the depth of this self-disclosure statements.
Evaluate the information of the self-disclosure statement:
    - Score 1: Statements that provide general or routine information only, without any personal reference
    - Score 2: Statements providing general information about the user
    - Score 3: Statements revealing personal information that exposes self or people close to the user, such as descriptions of physical appearance and behavior
Only give the score (number) without any explanation
\end{lstlisting}

\noindent\textbf{Thoughts.}
\begin{lstlisting}[style=llmprompt]
As a social scientist, your task is to annotate the depth of this self-disclosure statements.
Evaluate the thoughts of the self-disclosure statement:
    - Score 1: No indication of any thoughts or ideas on any subject that refer to the user personally; expressing of general ideas only
    - Score 2: Statements expressing the user's personal thoughts on past events or future plans
    - Score 3: Statements expressing thoughts relating to the user's personal characteristics, physical appearance, health, or intimate and wishful ideas
Only give the score (number) without any explanation
\end{lstlisting}

All the prompts and implementation are adapted directly from the previous study \cite{tran2025robot}.

\section{Additional Results}
\label{app:results}

\subsection{Baseline Comparability} \label{app:baseline_results}

We first tested whether the four well-being scores differ on pre-test. Separate HC3-robust Type III writing type $\times$ AI-Assistance Type ANOVAs revealed no significant effects for SWLS (gratitude writing type: $F(1,395)=0.24$, $p=0.627$; AI-assistance mode: $F(1,395)=0.02$, $p=0.886$; interaction: $F(1,395)=0.60$, $p=0.441$), GQ-6 ($F(1,395)=0.28$, $p=0.600$; $F(1,395)=0.47$, $p=0.494$; and $F(1,395)=0.01$, $p=0.921$, respectively), SPANE-B ($F(1,390)=1.04$, $p=0.308$; $F(1,390)=0.02$, $p=0.899$; and $F(1,390)=1.52$, $p=0.219$), or moral elevation ($F(1,395)=0.05$, $p=0.815$; $F(1,395)=0.06$, $p=0.811$; and $F(1,395)=0.42$, $p=0.518$). Thus, we detected \textbf{no statistically significant differences among the four well-being scores before the intervention}.

\subsubsection{Overall Pre--Post Changes}
We first examine changes in well-being across all four conditions using paired-samples $t$-tests. SWLS increases by 1.44 points, from $M=22.31$ to $M=23.75$, $t(383)=7.47$, $p<0.001$, $d_z=0.38$, 95\% CI $[1.06, 1.82]$. GQ-6 increases by 0.63 points, from $M=33.46$ to $M=34.09$, $t(378)=3.29$, $p=0.001$, $d_z=0.17$, 95\% CI $[0.25, 1.01]$. For affect balance, SPANE-B increases by 0.46 points, but this change is not significant, $t(375)=1.72$, $p=0.086$, $d_z=0.09$, 95\% CI $[-0.07, 0.99]$. Examining its components helps explain this result: SPANE-P increases by 1.59 points, from $M=20.99$ to $M=22.57$, $t(377)=8.89$, $p<0.001$, $d_z=0.46$, 95\% CI $[1.24, 1.94]$, while SPANE-N also increases by 1.11 points, from $M=10.53$ to $M=11.64$, $t(384)=6.01$, $p<0.001$, $d_z=0.31$, 95\% CI $[0.75, 1.47]$. Thus, the increase in positive affect is partly offset by a concurrent increase in negative affect. Finally, moral elevation increases by 0.71, from $M=5.79$ to $M=6.50$, $t(375)=12.16$, $p<0.001$, $d_z=0.63$, 95\% CI $[0.60, 0.83]$.

\end{document}